\documentclass[useAMS,usenatbib]{mn2e} 

\usepackage{times} 
\usepackage{color} 
\usepackage[dvipsnames]{xcolor}
\usepackage[normalem]{ulem} 
\usepackage{cancel}
\usepackage{amsmath}
\usepackage{amssymb} 
\usepackage{graphicx}

\newcommand{\be}{\begin{equation}}
\newcommand{\ee}{\end{equation}}

\newcommand{\ionMgii}{\mbox{Mg\,{\sc ii}}} 
\newcommand{\ionCiv}{\mbox{C\,{\sc iv}}}

\title[MHD Disc Winds and Line Width Distributions]{MHD Disc Winds and Line Width Distributions}
\author[L. S. Chajet and P. B. Hall]{L. S. Chajet$^{1}$ \thanks{E-mail: lchajet@yorku.ca} 
and P. B. Hall$^{1}$ \\
$^{1}$Department of Physics and Astronomy,
York University, Toronto, Ontario M3J 1P3, Canada} 

\begin{document}


\pagerange{\pageref{firstpage}--\pageref{lastpage}} \pubyear{2012}

\maketitle

\label{firstpage}

\begin{abstract}
We study AGN emission line profiles combining an improved version of the 
accretion disc-wind model of Murray \& Chiang  with the magneto-hydrodynamic 
model of  Emmering et al. 
We show how the shape, broadening and shift of the \ionCiv \,\! line depend 
not only on the viewing angle to the object but also on the wind launching
angle, especially for small launching angles. 
We have compared the dispersions in our model  \ionCiv \,\! linewidth 
distributions to observational upper limit on that dispersion, considering 
both smooth and clumpy torus models. 
As the torus half-opening angle (measured from the polar axis) increases 
above about 18\degr, increasingly larger wind launching angles are required 
to match the observational constraints.
Above a half-opening angle of about 47\degr, no wind launch angle (within 
the maximum allowed by the MHD solutions) can match the observations. 
Considering a model that replaces the torus by a warped disc yields the
same constraints obtained with the two other models.  
\end{abstract}

\begin{keywords} 
galaxies: active -- galaxies: nuclei -- (galaxies:) quasars: emission lines 
\end{keywords}

\section{Introduction}

Broad Emission Lines (BELs) are one characteristic feature of the spectra of 
Type 1 Active Galactic Nuclei (AGNs). 
Lines arising from high-ionization species are generally
blueshifted and single-peaked \citep[e.g.][]{Sulentic+95, Sulentic+00, Vanden Berk+01}. 
Quasars (and AGN in general) are powered by accreting mass into a central 
super-massive black hole (SMBH). The accreting mass is assumed to form a 
disc-like structure, responsible for most of the ultraviolet (UV) and optical continuum 
emission. 
This continuum emission illuminates and ionizes dense gas 
surrounding the central engine and forms the Broad Line Region (BLR), where the BELs 
originate. 
The black hole, disc and BLR are embedded in a dusty  toroidal structure that obscures some 
lines of sight to the nucleus \citep[e.g.,][]{Elitzur08}. 
In the model of \citet{LE10}, the torus is replaced by a warped disc.

Currently there is no consensus on the nature of the BLR and numerous
models have been developed to explain it. 
The spectroscopic characteristics of the BEL
can be explained by lines arising from either an approximately spherical distribution of 
discrete clouds, with no preferred velocity direction \citep[e.g.,][]{Kaspi&Netzer99} 
or at the base of a wind from an accretion disc  \citep[e.g.,][]{MCGV95, MC97, BKSB97}.  
In the cloud scenario, the BLR is described as composed of numerous optically-thick 
clouds that, photoionized by the continuum-source emission, are the emitting entities 
responsible for the observed lines. Although this model can explain many observed spectral 
features, it also leaves several unsolved issues, such as the formation and confinement 
of the clouds \citep[e.g.,][]{Netzer90}. 
The two relevant time-scales for these clouds are the sound crossing time $t_{\rmn {sc}}$ 
and the dynamical time $t_{\rmn {dyn}}$. According to the models, the masses of individual 
BLR clouds are below their Jeans mass, therefore without a confinement mechanism 
such clouds will disintegrate on a time-scale $t_{\rmn {sc}} \ll t_{\rmn {dyn}}$,  
in which case they would need to be continuously produced. 
In addition, the number of  clouds needed to reproduce the observed smoothness 
of BEL profiles \citep{Arav+98, DWCBN99} is implausibly large.  
Furthermore, even if the clouds are confined, cloud-cloud collisions would
destroy the clouds on a dynamical timescale \citep[e.g.][]{Mathews-Capriotti85},  
again requiring a high rate of cloud formation or injection.

One approach aimed at solving the discrete-cloud model difficulties was proposed 
by \citet*{EBS92}. In their model, the BLR is associated with disc-driven, hydromagnetic 
winds and  the lines are formed  by clouds which are  confined by the magnetic pressure. 
Low-ionization line profiles, (e.g., \ionMgii), and high-ionization line profiles 
(e.g., \ionCiv) are produced in the wind at different latitudes and radii.
Within that framework, the estimated values of parameters such as 
ionizing flux, electron density, cloud filling factor, column density, and velocity are in 
agreement with values for these quantities inferred from observations.  
\citet{EBS92} consider emission models with and without electron scattering and  
attempt to construct a typical \ionCiv \, emission profile. Different blueshifts and 
line asymmetries are obtained by varying model parameters. 

\citet{MCGV95} and later \citet{MC97, MC98} proposed a wind 
model motivated by the similarities between broad emission lines in AGNs and other 
astrophysical objects, such as cataclysmic variables, protostars and X-ray binaries. 
They made the assumption that the outflow is continuous instead of being composed of
discrete clouds and showed that such a continuous, optically thick, radiatively driven wind
launched from just above the accretion disc can account for both the single-peaked nature
of AGN emission lines and their blueshifts with respect to the AGN systemic redshift,
although not for the magnitude of these shifts.
In a accelerating wind, the wind opacity in a given direction depends on the velocity 
gradient in that direction. The larger radial gradient of velocity in a radially accelerating 
wind means that the opacity seen by radially-emitted photons will be lower than the opacity
seen by photons emitted in other directions.  Thus, photons will
tend to escape radially and the resultant emission lines are single-peaked. 
In addition, the model high-velocity component of the wind naturally explains the 
existence of blueshifted broad absorption lines seen in an optically selected 
subset (15-20$\%$) of the quasar population, known as broad absorption line quasars.

Here we combine an improved version of the \citet{MC97} 
model with the \citet{EBS92} model and analyse the dependence of the resulting 
emission line profiles on several parameters. 
The plan of the paper is as follows. In Section~\ref{sect: Modified Wind Model}  we 
review the \citet{MC97} disc-wind model and the modifications that we have 
introduced in a companion paper (Hall et al., in preparation).
In Section~\ref{sect: MHD wind}  we outline 
the basics of MHD winds and show how we combined the two models. 
In section ~\ref{sect: Q} we calculate a key quantity of the model, $Q$: 
the line-of-sight gradient of the line-of-sight velocity.  
The line profile results are presented in Section~\ref{sect: Res}. 
In Section~\ref{sec:FWHMs} we follow \citet{Fine+08} and study the predicted BEL width
distribution incorporating two models for the escape probability from the BLR, including
the clumpy torus model of Nenkova et al. (2008a, 2008b),
and use the results to obtain constraints on the torus parameters.  We analyse the warped
disc models of \citet{LE10} in a similar fashion in Section~\ref{sect:warped discs}.
We present our conclusions in Section~\ref{sec:Conclusions}.

\section[]{The Modified Wind Model}
 \label{sect: Modified Wind Model} 
 
In a companion paper (Hall et al. 2012, in preparation) we extend the disc-wind model of 
\citet[][MC97 hereafter]{MC97} to the case of non-negligible radial and vertical velocities.
The new treatment retains a number of factors neglected in MC97 and introduces
the `local inclination angle' to account for the different effective inclinations to the line
of sight of different portions of the emitting region.
Below we summarize these modifications. 

\begin{figure} 
\includegraphics[width=.47\textwidth]{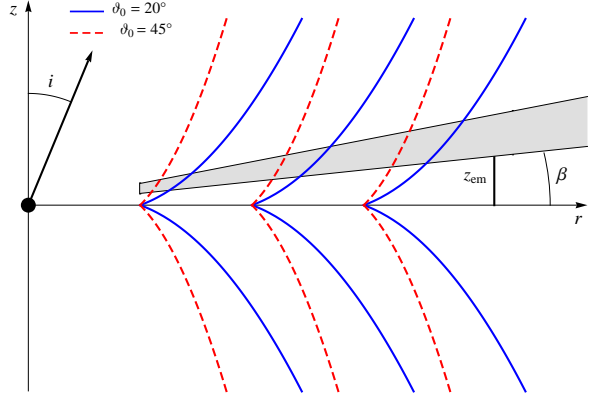}  
\caption{Streamlines for two different launching angles: $\vartheta_0=20\degr, 45\degr$. 
The blue line represents the base of the emitting region, tilted by an angle $\beta$ 
with respect to the disc plane.}
\label{Fig:Fig_0}
\end{figure}

As shown in Figure \ref{Fig:Fig_0}, we assume the SMBH is at the origin of a cylindrical 
coordinate system ($r, \phi, z$) with the $z$ axis normal to the accretion disc and the 
observer, in the $xz$ plane, making an angle $i$ with the disc axis. 
At any $r$, the azimuthally symmetric emitting region has its base at
$z_{\rmn {em}}=r\tan\beta(r)$ above the disc plane 
and has a density which drops off above the base as a Gaussian
with characteristic thickness that satisfies $l_{\rmn{em}}(r) \ll r$.
Because the photons originate in a narrow layer above the disk,
the emission region can be approximated as an emitting surface
with a source function $S_\nu$ that is a function of radius only.
The wind streamlines make an angle of $\vartheta(r,z)$ relative to the disc plane. 

Under these assumptions, the specific luminosity for a given line in the
direction of the observer, $L_\nu({\bf \hat{n}})$, is given by  
\begin{equation} \label{eq: define_L} 
L_\nu ({\bf \hat{n}}) = \int_{r_{\rmn {min}}}^{r_{\rmn {max}}} \!\!\!\!\!\!\!\!\!\!\!\!\! S_\nu(r) ~a(r) ~r ~dr 
 \int_0^{2\pi} \!\!\!\!\!\! [ 1-e^{-\tau_\nu(r,\phi,{\bf \hat{n}})} ] \cos \iota(r,\phi,{\bf \hat{n}}) ~d\phi 
\end{equation} 
where $I_\nu=S_\nu(r)[1-e^{-\tau_\nu}]$ is the specific intensity, $a(r)~r~dr~d\phi$ 
is the area of the emitting surface between cylindrical radii $r$ and $r+dr$, 
$\tau_\nu$ is the optical depth from $z_{\rmn {em}}(r)=r\tan\beta(r)$ to infinity along the 
direction $\bf{\hat{n}}$ from the location $(r,\phi)$, and $\iota(r,\phi,{\bf \hat{n}})$ is the 
local inclination angle between ${\bf \hat{n}}$ and the local normal to the surface at radius 
$r$ and azimuthal angle $\phi$.  For $i < 90 \degr - \beta(r)$, the area factor
and the local inclination angle are given by 
\begin{equation} \label{a(r)}
a(r) = \frac{1}{\cos\beta(r)}
\end{equation} 
and 
\begin{equation} \label{iota}
\cos \iota = \cos i \cos \beta(r) - \cos \phi \sin i \sin \beta(r) \, ,
\end{equation} 
respectively. 

To evaluate the optical depth $\tau_\nu$, MC97 expanded the projection of the wind 
velocity along the line of sight, $\mathbf{\vec{v}\cdot\hat{n}}$, in terms of $z-z_{\rmn {em}}$ 
to first order to obtain (their equation 12)

\begin{align} 
 \label{eq: e_vdotn}
\mathbf{\hat{n} \cdot \vec{v}}(r,\phi,z) 
& \cong \mathbf{\hat{n} \cdot \vec{v}}(r,\phi,z_{\rmn {em}}) 
+  \mathbf{\hat{n} \cdot \Lambda \cdot \hat{n}}
~\frac{(z-z_{\rmn {em}})}{\cos \iota/\cos\beta}   \nonumber  \\ 
& 
\equiv v_{\rmn {D}}(r,\phi,z_{\rmn {em}}) + 
\frac{(z-z_{\rmn {em}})}{l_{\rmn {em}}}
~v_{\rmn {sh}}(r,\phi,z_{\rmn {em}}).
\end{align} 

The zeroth order term in Equation \ref{eq: e_vdotn} is the Doppler velocity $v_{\rmn {D}}$: 
\begin{equation}
 \label{vD} 
 v_{\rmn {D}}=-v_{\phi} \sin\phi \sin i
+v_{\rmn {p}} \left(\cos\phi \cos\vartheta \sin i + \sin\vartheta \cos i \right) 
\end{equation} 
where $v_{\phi}$ and $v_{\rmn {p}}$ 
are, respectively, the azimuthal and poloidal velocities of the wind 
at $z = z_{\rmn {em}}$.  

The first order term in Equation \ref{eq: e_vdotn} involves the shear velocity 
$v_{\rmn {sh}}$, defined as $v_{\rmn {sh}}=l_{\rmn {em}}Q\cos\beta/\cos\iota$ 
where $Q$ \citep{RH78, RH83} is the line-of-sight gradient of the line-of-sight 
wind velocity:
\begin{equation} 
\label{eq:Q}
Q\equiv\mathbf{\hat{n} \cdot \Lambda \cdot \hat{n}}
\end{equation} 
where $\mathbf{\hat{n}}$ is the unit vector in the line of sight direction and $\mathbf{\Lambda}$ 
is the strain tensor.  
The entries of the strain tensor $\Lambda$ consist of spatial derivatives of velocity components. 
It is symmetric ($\Lambda_{ij} = \Lambda_{ji}$) and its elements are 
given in cylindrical coordinates by \citep[see e.g.][]{Batchelor67}:  
%
\begin{align}
\Lambda_{r \phi}  & = \frac{1}{2} \left( \frac{1}{r} \frac{\partial v_r}{\partial \phi} -
\frac{v_\phi}{r} + \frac{\partial v_{\phi}}{\partial r} \right),  \quad
\Lambda_{r z} = \frac{1}{2} \left( \frac{\partial v_r}{\partial z} +
\frac{\partial v_z}{\partial r} \right),   \nonumber  \\ 
\Lambda_{\phi z} & = \frac{1}{2} \left(\frac{\partial v_\phi}{\partial z} +
\frac{1}{r} \frac{\partial v_z}{\partial \phi} \right), \nonumber \\
%
\Lambda_{r r}  & =  \frac{\partial v_r}{\partial r},  \quad  \Lambda_{\phi \phi} =
\frac{1}{r} \frac{\partial v_\phi}{\partial \phi} + \frac{v_r}{r},  \quad    \Lambda_{z z}  =
 \frac{\partial v_z}{\partial z}.
\end{align}
In terms of these $\Lambda_{ji}$, the quantity $Q$ is: 
\begin{equation} 
\label{eq:Q-using-Lambda}
\begin{split}
Q & = \sin^2 i \left[\Lambda_{r r}  \cos^2 \phi + \Lambda_{\phi \phi} \sin^2 \phi -
2 \Lambda_{r \phi} \sin \phi  \cos \phi \right] +  \\
& \quad \cos i \left[2 \Lambda_{r z} \sin i \cos \phi  +
\Lambda_{z z}  \cos i - 2 \Lambda_{\phi z} \sin i \sin \phi \right] \\
\end{split} 
\end{equation} 
Assuming azimuthal symmetry, all the $\partial / \partial \phi = 0$ and the simplified
expressions for the different $\Lambda_{ij}$ are:
%
\begin{align}
\label{eq:epsilon1}
\Lambda_{r \phi} & = \frac{1}{2} \left( \frac{\partial v_{\phi}}{\partial r}
-\frac{v_\phi}{r}   \right),  \quad
\Lambda_{r z} = \frac{1}{2} \left( \frac{\partial v_r}{\partial z} +
\frac{\partial v_z}{\partial r} \right),  \nonumber \\  
\Lambda_{\phi z} & = \frac{1}{2} \frac{\partial v_\phi}{\partial z}, \quad
%
\Lambda_{r r}  =  \frac{\partial v_r}{\partial r},  \quad  \Lambda_{\phi \phi} =
\frac{v_r}{r},  \quad    \Lambda_{z z}  =  \frac{\partial v_z}{\partial z}
\end{align}
In the above, the novel element introduced in Hall et al. (2012, in preparation) 
is the dropping of the assumption of $v_r \ll v_{\phi}$, thus allowing for 
non-negligible radial and vertical velocities.  

Including that and several factors that have been omitted or considered negligible 
in the original MC97 work, the final expression for the specific luminosity in a line of 
central frequency $\nu_0$ emitted from a disc with a disc wind is given by 
\begin{equation} 
\begin{split}
\label{Lnu}
L_\nu (i) & = \int_{r_{\rmn {min}}}^{r_{\rmn {max}}} S_\nu(r)~a(r)~r ~dr 
\int_0^{2\pi} \cos\iota(r,\phi,i)  ~ \times \\
&  (1-\exp[-\tau(r,\phi,i) \times e_\nu(r,\phi,i) \times e^{-x_\nu^2(r,\phi,i)}]) ~d\phi 
\end{split}
\end{equation}
where 
\begin{equation} 
\tau(r,\phi,i)  \equiv  
\frac{ck_0(r)/2\nu_0}{\sqrt{Q^2(r,\phi,i)+q_{\rmn {tt}}^2(r,\phi,i)}}  
\label{eq: tau} 
\end{equation}
\begin{equation}
e_\nu(r,\phi,i) \equiv
{\rmn {\, erfc}}
\left( 
-\frac{\nu-\nu_{\rmn {D}}(r,\phi,i)}
{\sqrt{2}\Delta \nu_{\rmn {tt}} \sqrt{1+q_{\rmn {tt}}^2(r,\phi,i)/Q^2(r,\phi,i)}} 
\right)\\
\label{eq: enu} 
\end{equation}
\begin{equation}
x_\nu^2(r,\phi,i) \equiv \frac{1}{2} \left(  
\frac{\nu-\nu_{\rmn {D}}(r,\phi,i)}{
\Delta \nu_{\rmn {tt}} \sqrt{1+Q^2(r,\phi,i)/q^2_{\rmn {tt}}(r,\phi,i)}}
\right)^2 
\label{eq: x2}  
\end{equation} 
and erfc is the complementary error function.  
In the above expression, the $k_0(r)$ is the integrated line opacity (units of Hz/cm) 
at $z_{\rmn {em}}$, and we have defined the Doppler-shifted central frequency of 
the line emitted towards the observer from location ($r,\phi$) on the emitting surface,
$\nu_{\rmn {D}} = \nu_0 (1+ v_{\rmn {D}}/c)$, and the `thermal $Q$', the ratio of the 
characteristic thermal plus turbulent velocity 
of the ion to the thickness of the emitting layer along the line of sight, 
$q_{\rmn {tt}}(r,\phi,i)=v_{\rmn {tt}}\cos\iota(r,\phi,i)/l_{\rmn {em}}(r)\cos\beta(r)$, with 
$ v_{\rmn {tt}}^2 \equiv v_{\rmn {th}}^2 + v_{\rmn {turb}}^2 $. 
The effective frequency dispersion of the line is given by 
$\Delta \nu_{\rmn {tt}}=\nu_0 v_{\rmn {tt}}/c$.
The $z$-dependent quantities are evaluated at $z=z_{\rmn {em}}$ where applicable.
The emission region thickness is given by
\begin{equation} \label{define_lem}
l_{\rmn {em}}(r)=0.1{z_{\rmn {em}}} \left[{v_{\rmn {tt}}+
v_{\rmn {p}}(r,z_{\rmn {em}}) \over v_{\rmn {tt}}+v_\infty(r,z_{\rmn {em}})}\right].  
\end{equation}

\section{Magneto-hydrodynamic wind model}
 \label{sect: MHD wind} 

The presence of an ordered magnetic field threading an AGN accretion disc has 
been suggested by several authors \citep[e.g.,][]{BP82, CL94, KK94} as a 
mechanism able to either confine the clouds \citep[as in][]{EBS92} or direct the 
outflow velocity field \citep[e.g.,][]{Everett05}.  
Along with most works in the field, we do not discuss here the origin of the magnetic 
field, but assume it is present and study its effects within the postulated framework. 

The standard magneto-hydrodynamic (MHD) wind equations are:
\begin{subequations} \label{eq: MHD-eqs}
\begin{align}
\frac{\partial \rho}{\partial t} +
\mathbf{\nabla} \cdot (\rho \mathbf{v}) & =  0 \\ 
\rho \frac{\partial \mathbf{v}}{\partial t} + \rho (\mathbf{v} \cdot \mathbf{\nabla}) \mathbf{v} & = 
-\mathbf{\nabla}p - \rho \mathbf{\nabla} \Phi_{\rmn {g}} +  \frac{1}{4 \pi} (\mathbf{\nabla} \times
\mathbf{B} ) \times \mathbf{B}  \label{eq:Mom Equa} \\
\frac{\partial \mathbf{B}}{\partial t} & =  \mathbf{\nabla} \times (\mathbf{v}  \times \mathbf{B} )\\ 
\mathbf{\nabla} \cdot \mathbf{B} & =  0,
\end{align}
\end{subequations}
where $\mathbf{B}$ and $\mathbf{v}$ are respectively the magnetic and
velocity fields, $\rho$ is the mass density, $p$ is the thermal pressure and
$\Phi_{\rmn {g}}$ is the gravitational potential. 

We look for steady state wind solutions for this model. In that case (i.e. when 
$\partial/\partial t = 0$), there are 
conserved quantities along each magnetic field line \citep[e.g.,][]{Mestel68}, such as
\begin{description}
  \item  mass to magnetic flux ratio, 
\begin{equation} \label{eq:k}
\frac{k}{4 \pi} = \frac{\rho v_{\rmn {p}}}{B_{\rmn {p}}}
\end{equation} 
  \item specific angular momentum, 
  \begin{equation} \label{eq:l}
l = r \left(v_{\phi} -
\frac{B_\phi}{k} \right) \\
\end{equation} 
  \item specific energy 
\begin{equation} \label{eq:e}
e =\frac{v^2}{2} + h + \Phi_{\rmn {g}} -
 \frac{ r  \Omega B_{\phi}  }{k}
\end{equation} 
\end{description}
where $h$ is the specific enthalpy, and, owing to the axisymmetry of the problem, we have 
separated the velocity and magnetic fields into their poloidal and azimuthal components: 
 \begin{equation}
 \mathbf{v} =  \mathbf{v}_p +  \mathbf{v}_{\phi} \; \; \qquad \qquad
 \hfill
  \mathbf{B} =  \mathbf{B}_p +  \mathbf{B}_{\phi}
\end{equation}

\subsection{Self-similar solutions} 
 \label{subsect: self-sim sol.} 

Solutions of the steady, axisymmetric, non-relativistic ideal MHD equations assuming 
a spherically self-similar scaling were obtained by e.g. \citet[][BP82 hereafter]{BP82} 
for the cold plasma outflow from the surface of a Keplerian disc. 
This solution for the field can be written in terms of variables $\chi$, $\xi(\chi)$, $\phi$, 
and $r_0$,  which are related to the cylindrical coordinates via
\begin{equation}\label{eq:r}
\mathbf{r} \equiv \left[r, \phi, z\right] = \left[r_0 \xi(\chi), \phi, r_0 \chi \right],
\end{equation}
where the adopted independent variables $(r_0, \chi)$ are a pair of spatial coordinates 
analogous to $(r, z)$.
The function $\xi(\chi)$ describes the shape of the field lines and, in the general case, is not 
\textit{a priori} known, but found as part of a self-consistent solution to the MHD equations. 
The flow velocity components are given by
\begin{equation}\label{eq:v}
\mathbf{v} = \left[\xi'(\chi)f(\chi), g(\chi), f(\chi) \right] \sqrt{\frac{G M}{r_0}},
\end{equation}
where a prime denotes differentiation with respect to $\chi$, and $G$ and $M$ 
are respectively the gravitational constant and the mass of the central black hole.  

In this self-similar model, the scaling of the speed $v$, magnetic field amplitude $B$, 
and gas density $\rho$ with the spherical radial coordinate $r$ is determined from 
the relation $B/\sqrt{\rho}  \propto r^{-1/2}$ and from the assumption that 
$r_0^2 \rho v_0$ is independent of $r_0$, from where $\rho_0 \propto r_0^{-3/2}$ 
and $B_0 \propto r_0^{-5/4}$. 
Other authors \citep[e.g.,][]{CL94, EBS92} generalized this class of self-similar 
solutions by considering winds with a density scaling $\rho  \propto r^{-b}$, for 
which $B \propto r^{-(b+1)/2}$. 
Note that, in this context,  the BP82 solution corresponds to $b = 3/2$. The magnetic 
field and density at  arbitrary positions can be then written, in accordance with the 
self-similarity {\it Ansatz} (\ref{eq:r}), as ${\bf{ B} }= B_0(r_0) {\bf{b}}(\chi)$ and 
$\rho = \rho_0 (r_0) \varrho(\chi)$.
On the disc plane the rotational velocity, $v_{\phi}$, is Keplerian and scales as 
$v_{\phi} \propto r_0^{-1/2}$. The functions $\xi(\chi)$, $f(\chi)$ and $g(\chi)$ have 
to satisfy the flow MHD equations subject to the above scalings of $\rho$, $B$, and 
$v_{\phi}$ and boundary conditions. In particular, at the disc surface $\xi(0)=1$, 
$f(0)=0$ and $g(0)=1$.

Following BP82, we introduce the dimensionless expressions of the integrals of motion
defined in equations \ref{eq:k}-\ref{eq:e}, in terms of which the solutions are defined:
\begin{equation} \label{eq:kappa}
\kappa = k(1+\xi'_0)^{1/2}\frac{(GM/r_0)^{1/2}}{B_{0}}
\end{equation}
\begin{equation} \label{eq:lambda}
\lambda = \frac{l}{(GMr_0)^{1/2}} \\
\end{equation}
\begin{equation} \label{eq:epsilon}
\epsilon = \frac{e}{(GM/r_0)}
\end{equation} 
The parameters of the model are $\epsilon$, $\lambda$ and $\kappa$ and  $\xi'_0$. 
However, due to the regularity conditions that must be satisfied, these parameters 
are not independent.
Combining equations \ref{eq:lambda} and \ref{eq:epsilon} gives 
$\epsilon~=~\lambda - \frac{3}{2}$.
The value of $\xi'_0\equiv \xi'(\chi=0)$ 
must be chosen to ensure the regularity of the solution at the Alfv\'en point. 
\footnote{
The Alfv\'en point is where the poloidal velocity of the fluid  is equal to the poloidal 
component $v_{\rmn {pA}}$ of the Alfv\'en velocity $v_{\rmn {A}}$ defined in Eq. \ref{eq:Alfven vel.}
} 
The solutions are therefore parametrized only by two numbers, which can be chosen 
to be $\kappa$ and $\lambda$ (e.g. BP82). 

The Alfv\'en speed,  $v_{\rmn {A}}$,  is the characteristic velocity of the propagation of magnetic 
signals in an MHD fluid and is defined by: 
\begin{equation} \label{eq:Alfven vel.}
v_{\rmn {A}} = \frac{B^2}{\mu_0 \, \rho}, 
\end{equation}
where $\mu_0$ is the vacuum permeability. 
Another important characteristic quantity in magnetized fluids is the Alfv\'en Mach number at 
each position. The square of this quantity is expressed in the present model as 
\begin{equation} \label{eq:m}
m(\chi) = \frac{v_{\rmn {p}}^2}{v_{\rmn {pA}}^2} = 
 \frac{4 \pi \rho \, v_{\rmn {p}}^2}{B_{\rmn {p}}^2} = \kappa \, f(\chi) \, \xi(\chi) \, J(\chi),
\end{equation}
where
\begin{equation} \label{eq:J}
J(\chi) = \xi(\chi) - \chi \;\xi'(\chi)
\end{equation} 
is the determinant of the Jacobian matrix of the transformation
$(r, z) \rightarrow (r_0, \chi)$ and $v_{\rmn {pA}}$ is the 
poloidal component of the Alfv\'en velocity. 

The function $g(\chi)$ can be expressed in terms of the function $m$ and the specific
angular momentum, $\lambda$: 
\begin{equation} \label{eq:g}
    g(\chi) = \frac{\xi^2(\chi) - \lambda \, m(\chi)}{\xi(\chi)\left[ 1 - m(\chi) \right]}. 
\end{equation} 
From this expression we can see that the point corresponding to $m = 1$ is a singular 
point of the problem. In particular, to avoid unphysical solutions there, we must  have 
$\xi_{\rmn {A}} = \xi(\chi_{\rmn {A}})=\lambda^{1/2}$, where the subscript A refers to the 
Alfv\'en point. 

Expressing $f$ and $g$ by Eqs (\ref{eq:m}) and (\ref{eq:g})  in terms of the Alfv\'enic 
Mach number and of the function $\xi(\chi)$ and its derivatives, Eq. (\ref{eq:e}) is 
transformed into a fourth degree equation for the function $f(\chi)$:
\begin{equation} \label{eq:BP82 Eq 2.12}
T- f^2 (1+ \xi'^2) =\left[ \frac{(\lambda-\xi^2) \, m}{\xi(1-m)} \right]^2,
\end{equation}
where
\begin{equation}
T = \xi^2 +  \frac{2}{\sqrt{\xi^2+\chi^2}} - 3.
\end{equation}
Using the differential form of equation \ref{eq:BP82 Eq 2.12} combined with the $z$-component 
of the momentum equation (Eq.  \ref{eq:Mom Equa}), BP82 obtain a second-order differential 
equation for $\xi(\chi)$. 
The flow is then fully specified by that equation and equation \ref{eq:BP82 Eq 2.12}, plus the 
boundary conditions, $\xi(0)=1$ and $\xi'(0)=\xi'_0$.  

The model of \citet[][hereafter EBS92]{EBS92} represents a simplified version of BP82 solution. 
The EBS92 solution corresponds to the case in which the solution asymptotically approaches 
$n =1$ as $\chi \rightarrow \infty$, where $n$ is the square of the Mach number for the fast 
magnetosonic mode for an arbitrary scaling of density  $\rho_0 \propto r_0^{-b}$ and magnetic 
field $B_0 \propto r_0^{-(b+1)/2}$.
The quantity $n$ is given by (e.g. BP82, BPS92): 
\begin{equation} \label{eq:n}
n = \frac{4 \pi \rho \, v_{\rmn {p}}^2}{B^2} = \frac{ \kappa \, \xi \,  f^3J  (1 + \xi'^2)}{T}. 
\end{equation}

While BP82 found their solutions by integrating a second-order differential equation, 
EBS92 impose {\it a priori} the functional form of the solution so that it will asymptotically 
tend to the BP82 solution.      
In their equation (3.19), EBS92 give an explicit form for the function $\xi(\chi)$:
\begin{equation} \label{eq:xi}
    \xi = \left( \frac{\chi}{c_2} +  1\right)^{1/2}, 
\end{equation}
where $c_2 = \frac{1}{2}\tan \vartheta_0$ was chosen to ensure that the field lines
make an initial angle $\vartheta_0$ with the disc plane, so that
$\cot \vartheta_0 = \xi'_0$ 
and the subscript $0$ means that the quantities are evaluated at the disc plane.
It can be demonstrated  \citep[e.g. BP82,][]{Heyvaerts96} that there is an upper 
limit for this angle: $\vartheta_0 < 60\degr$.

EBS92 found that to satisfy the condition that $n \rightarrow 1$ when $\chi \rightarrow \infty$, 
the parameters $\kappa$ and $\lambda$ must be related by:
\begin{equation} \label{eq: kappa_lambda}
\kappa = 2 \left(\frac{3}{2 \lambda-3}\right)^{3/2}
\end{equation}
In addition, the asymptotic value of the function $f$ is given by 
\begin{equation}
f_{\infty} =  \left(\frac{2 \lambda-3}{3}\right)^{1/2}.
\end{equation} 
Thus, in this model the solutions depend on $\lambda$ and $\vartheta_0$.
Figure \ref{Fig:Fig_0} depicts the streamlines for two different launching angles. 

Note that in the general case, in order to find the flow variables, we should have solved a 
second-order differential equation $ \xi''=\xi''(\chi, \xi, \xi', f(\chi))$, with $f(\chi)$ given 
implicitly by equation \ref{eq:BP82 Eq 2.12}. 
However, by using the EBS92 model we could evaluate $f'$, $m$ and $g'$ from an 
analytic estimate for $f$. The point of having used an analytical functional form for 
$f(\chi)$ has to do with the inclusion of the velocity field in the MC97 model. 
The opacity depends on the projection of the line of sight (LOS) component of the 
gradient of the LOS velocity through the quantity $Q$ (equation \ref{eq:Q-using-Lambda}), 
which involves the spatial derivatives of velocity components. 
Combining the EBS92 functional form for $\xi(\chi)$ and $\xi_{\rmn {A}} = \lambda^{0.5}$ it 
is straightforward to obtain $\chi_{\rmn {A}} = (\lambda-1) c_2$. 
In the general case this quantity must be found numerically as part of the solution. 
However, in the adopted framework, all related quantities at the Alfv\'en point are easily 
found (because $m_{\rmn {A}} \equiv m(\chi_{\rmn {A}}) = 1$
).  In particular, 
\begin{equation}
f_{\rmn {A}} = \frac{1}{\kappa \sqrt{\lambda} J_{\rmn {A}}} = 
\frac{1}{\kappa\sqrt{\lambda}[\xi (\chi_{\rmn {A}}) - \chi_{\rmn {A}} \xi'(\chi_{\rmn {A}})]}. 
\end{equation} 

The derivative of $f(\chi)$ at $\chi=0$, $f'(\chi=0)=f'_0$, is given by BP82 (their equation 2.23c), 
reproduced here: 
\begin{equation}
f'_0=  \frac{(3\xi'^{2}_0-1)^{1/2} }{\kappa^2 \left[(\lambda-1)^2+(1+\xi'^{2}_0)\right]^{1/2}}\;.
\end{equation} 
We thus adopt
\begin{equation}
f(\chi)= f_\infty  \frac{ e^{k_1 \chi} - 1}{e^{k_1 \chi} + k_2} \;
\end{equation} 
and look for $k_1$ and $k_2$ such that the conditions for $f_{\rmn {A}}$ and 
$f'_0$ are satisfied.  
Once $f(\chi)$ is found, $m(\chi)$ and thus $g(\chi)$ are
obtained. We then have the three wind velocity components expressed in analytical form. 

Note that, as already mentioned, EBS92 model postulates that the emission lines 
arise in clouds confined by an MHD flow. However,  we follow MC97 and 
\citet[][MC98 hereafter]{MC98} 
in assuming that the lines form in a continuous medium. 
As will be discussed in section \ref{sect: Res}, we consider line emissivity obtained 
by CLOUDY photoionization model, different from either of the emissivity laws 
adopted by EBS92. Two of those emissivity models include electron scattering, 
which is not considered in our model.   
In EBS92 the dimensionless angular momentum $\lambda$ and the launch angle 
are fixed, while in our work the former is still fixed but the latter is varied to study 
its effect on the profiles.   
It is important to note that, while EBS92 obtains the line luminosity integrating in the 
two poloidal variables, we include the $z$-integral in the optical depth expression.

\section{Determination of $Q$  for self-similar MHD winds}
\label{sect: Q}
For self-similar solutions of MHD winds, the derivatives needed to obtain the
different $\Lambda_{ij} $ that appear in the quantity $Q$ have to be evaluated
using the rules for changing variables \citep[e.g.,][]{Konigl89} 
\begin{equation} \label{eq:d_dr}
 \frac{\partial}{\partial r} =  \frac{1}{J} \frac{\partial }{\partial r_0}  -
  \frac{\chi}{r_0 J} \frac{\partial}{\partial \chi},  
\end{equation}
\begin{equation} \label{eq:d_dz}
 \frac{\partial}{\partial z} =  -\frac{\xi'}{J} \frac{\partial }{\partial r_0} +
  \frac{\xi}{r_0 J} \frac{\partial}{\partial \chi}, 
\end{equation} 
where $J(\chi)$ has been defined in Eq. (\ref{eq:J}). 
Thus, we have the following expressions, where for clarity we omit the functional dependence 
of the dependent variables: 
\noindent
{\begin{equation}
\Lambda_{rr} = -\frac{1}{J} \sqrt{\frac{GM}{r_0^3}} \;\left( \frac{\xi' f}{2} + \chi \left( \xi'' f + \xi' f'\right) \right)
\end{equation}
 } 
\noindent
{\begin{equation}
\Lambda_{\phi\phi} = \sqrt{\frac{GM}{r_0^3}} \; \frac{\xi' f}{\xi}
\end{equation} 
} 
\noindent
{\begin{equation}
\Lambda_{zz} = \frac{1}{J} \sqrt{\frac{GM}{r_0^3}} \; \left( \frac{\xi' f}{2} + \xi f'\right)
\end{equation} 
} 
\noindent
{\begin{equation}
\Lambda_{rz} = \frac{1}{2J} \sqrt{\frac{GM}{r_0^3}}
\left[ \left(\frac{\xi'^2}{2} +
 \xi \xi'' -\frac{1}{2} \right) \,\! f + \left(\xi \xi' - \chi \right) \, \! f'  
 \right]
\end{equation} 
}
\noindent
{\begin{equation}
\Lambda_{r \phi} = \frac{1}{2} \; \sqrt{\frac{GM}{r_0^3}}  
\left[ -\frac{1}{J} \left(\frac{g}{2} + \chi g' \right)  + \frac{g}{\xi}   \right]
\end{equation} 
}
\noindent
{\begin{equation}
 \Lambda_{\phi z} = \frac{1}{2J} \; \sqrt{\frac{GM}{r_0^3}} \left( \frac{\xi' g}{2} + \xi g'\right)
\end{equation} 
}

For the particular form of $\xi(\chi)$ given by EBS92 the corresponding expressions for the 
strain tensor entries are:
\noindent
{\begin{equation}
\Lambda_{rr} = - \sqrt{\frac{GM}{r_0^3}} \;
\left[ \frac{f c_2 + 2 \chi^2 f' + 2 \chi f' c_2}{2\left(\chi + c_2 \right) \left( \chi + 2 c_2 \right)} \right]
\end{equation} 
}
\noindent
{\begin{equation}
\Lambda_{\phi\phi} = \sqrt{\frac{GM}{r_0^3}} \;\left[  \frac{f}{2 (\chi + c_2)} \right]
\end{equation} 
} 
\noindent
{\begin{equation}
\Lambda_{zz} = \sqrt{\frac{GM}{r_0^3}} \;
\left[ \frac{f + 4 f' \left(\chi + c_2 \right) }{2 (\chi + 2 c_2)} \right]
\end{equation} 
} 
\noindent
{\begin{equation}
\Lambda_{rz}\!=\!\sqrt{\frac{GM}{r_0^3}} \! \left[
\frac{
4 \left( \chi + c_2 - 2 c_2 \chi^2 - 2 c_2^2 \chi \right) \!\! f'
\!-\!\left(1 + 4c_2 \chi + 4 c_2^2\right)f 
} 
{8 \sqrt{c_2} \left(\chi + 2c_2 \right) \sqrt{\chi + c_2}}
\right]
\end{equation} 
}
\noindent
{\begin{equation}
\Lambda_{r \phi} = -\sqrt{\frac{GM}{r_0^3}} \sqrt{c_2}\left[
\frac{ (2 \chi + 3 c_2) g +  2 \chi ( \chi + c_2) g'} {2 \left(\chi + 2 c_2 \right) \sqrt{\chi + c_2 }}
\right]
\end{equation} 
}
\noindent
{\begin{equation}
\Lambda_{\phi z} = -\sqrt{\frac{GM}{r_0^3}} \left[
\frac{ g + 4  g' \chi + 4 g' c_2 } {4 \left(\chi + 2 c_2 \right) }
\right]
\end{equation} 
}

\section{Line Profiles} 
 \label{sect: Res} 

We evaluated the line luminosity (Eq. \ref{Lnu}) using the EBS92 solution to estimate the 
quantities included there and in the associated equations \ref{eq: tau} and \ref{eq: x2}.  
In summary, $v_{\rmn {D}}$ and $Q$ are computed as functions of position ($r,\phi,z$) 
from the velocity field given by the EBS92 model.  Then, these two quantities 
and the `thermal $Q$', $q_{\rmn {tt}}$, 
are used to evaluate the optical depth 
$\tau(r,\phi,i)$ (Eq. \ref{eq: tau}) and the quantities
 $e_\nu(r,\phi,i)$ (Eq. \ref{eq: enu})
 $x_\nu(r,\phi,i)$ (Eq. \ref{eq: x2}).
We emphasize again that the integral in the $z$ direction is included in the optical 
depth expression. 
We then calculate $L_\nu ({\bf \hat{n}})$ by integrating over all ($r,\phi$) (Eq. \ref{Lnu}). 
The process is repeated for different $\nu$ values to build up the profile of the given 
emission line for the given input parameters.

We have computed the \ionCiv \, line profile for different combinations 
of inclination angle, $i$ and initial angle, $\vartheta_0$.  
We also studied the results of changing the initial density and the exponent of the power 
law that governs the radial behaviour of the density.
The specific luminosity from each component of the \ionCiv \, doublet 
is computed separately, and then the results added together. 
In Table \ref{Tab: Table1} we list the meaning and adopted values of the main parameters in
the model. 
The fiducial values adopted for the density, density power-law exponent and thermal plus 
turbulent velocity are $n_0=10^{11}$ cm$^{-3}$, $b=2$ and $v_{\rm tt}=10^7$ cm s$^{-1}$,  
respectively. 

We determine the source function for our simulations by applying the 
reverberation mapping results of \citet{Kaspi+07}  to the radial line 
luminosity function $L(r)$ calculated by MC98 for a quasar with 
$L_{1350}\equiv \nu L_\nu(1350~{\rmn {\AA}}) =10^{46}$\,erg\,s$^{-1}$ 
and shown in their Figure 5b. 
According to that figure, the peak \ionCiv \, emission is reached 
at $R_{CIV}=10^{18}$\,cm, but the \citet{Kaspi+07} results show that 
$R_{CIV}$ is smaller for a quasar of that luminosity. 
Their Equation 3 gives
\begin{equation}\label{RCIV}
R_{CIV} = 6.216\times 10^{15} {\rm ~cm} 
\left( {L_{1350} \over 10^{43} {\rm ~erg\,s}^{-1} }\right)^{\alpha}, 
\end{equation}
where $\alpha = 0.55\pm 0.04$ in the original formulation and for simplicity we 
have adopted $\alpha = 0.5$.  
Eq.~\ref{RCIV} gives  $R_{CIV}=2\times 10^{17}$\,cm for a quasar with 
$L_{1350}=10^{46}$\,erg\,s$^{-1}$. 
We therefore empirically adjust all the radii in the MC98 Figure 5b 
line luminosity function down by a factor of five. 
\footnote{
The reader should be cautioned that, strictly speaking, 
this translation of the line-continuum lag measured in reverberation mapping 
experiments and the peak of the radial emissivity distribution of the line is not 
straightforward in the general case. 
}
Each point in the line luminosity function now gives the line luminosity 
$L(r_i)$ in a logarithmic bin spanning a factor of $\sqrt{10}$ in radius 
centred on adjusted radius $r_i$ for a quasar with 
$L_{1350}=10^{46}$\,erg\,s$^{-1}$.

\begin{table} 
\caption{Set of parameters used in the simulation.}
\begin{tabular}{@{}lll} 
\hline
Variable &   Value  &    Explanation\\
\hline
$M_{\rmn {BH}}$ &   10$^8$ $M_{\sun}$ &   Black hole mass\\
$L_{\rmn {UV}}$  &  10$^{46}$ erg s$^{-1}$  &  Quasar ionizing luminosity\\ 
$S_\nu(r)$  &  CLOUDY results &  Source function\\ 
$r_{\rmn {min}}$  & $2\sqrt{ { L_{\rmn {UV}} \over 10^{46} } \text{erg s}^{-1}} \times 10^{15}$ cm  
&  Inner BELR radius.\\ 
$r_{\rmn {max}}$  &  $2\sqrt{ { L_{\rmn {UV}} \over 10^{46} } \text{erg s}^{-1}} \times 10^{19}$ cm  
&  Outer BELR radius.\\     
$n_0$  &  $10^{7}-10^{13}$ cm$^{-3}$  &  Hydrogen number density at\\
& & $r_{\rmn {min}}$, declining as $r^{-b}$ thereafter\\
$b$ & 0.5, 1, 2 & Exponent in $n(r)\propto r^{-b}$\\ 
$i$   &  $5\degr-84\degr$  &  Observer inclination angle\\
$\vartheta_0$  &  $5\degr$, $10\degr$, $15\degr$, $30\degr$, $45\degr$, $57\degr$  & 
Streamline launch angle\\ 
$\tan\beta(r)$  &  0.1051  &  $z_{\rm em}= r\tan\beta(r)= r\tan(6\degr)$\\
$v_{\rmn {tt}}$  &  $10^6-10^7$ cm s$^{-1}$  &  Thermal+turbulent speed of ion\\
$\chi_i$  &  solar  &  Abundance of element\\
$\eta_i$  &  1  &  Ionization fraction of ion\\
$\lambda$  &  10 &  
Specific angular momentum\\ 
\hline
\end{tabular}
\label{Tab: Table1}
\end{table}

\subsection{Different $ \bf{v_{\rm \textbf{th}}} $}
\label{sec: subsec5.1} 

For a thermal velocity of $v_{\rmn {th}} = 10^{6}$ cm s$^{-1}$ and no turbulence,
the profiles of the individual components 
of the doublet are very narrow (FWHM $<$  inter-component separation) for small inclination 
angles. As a result, the combined profiles are double- (or multiple-) peaked. The effect is less 
pronounced for higher ($\gtrsim 45\degr$) $i$ values.

\begin{figure*}
\includegraphics[width=.825\textwidth,angle=0]{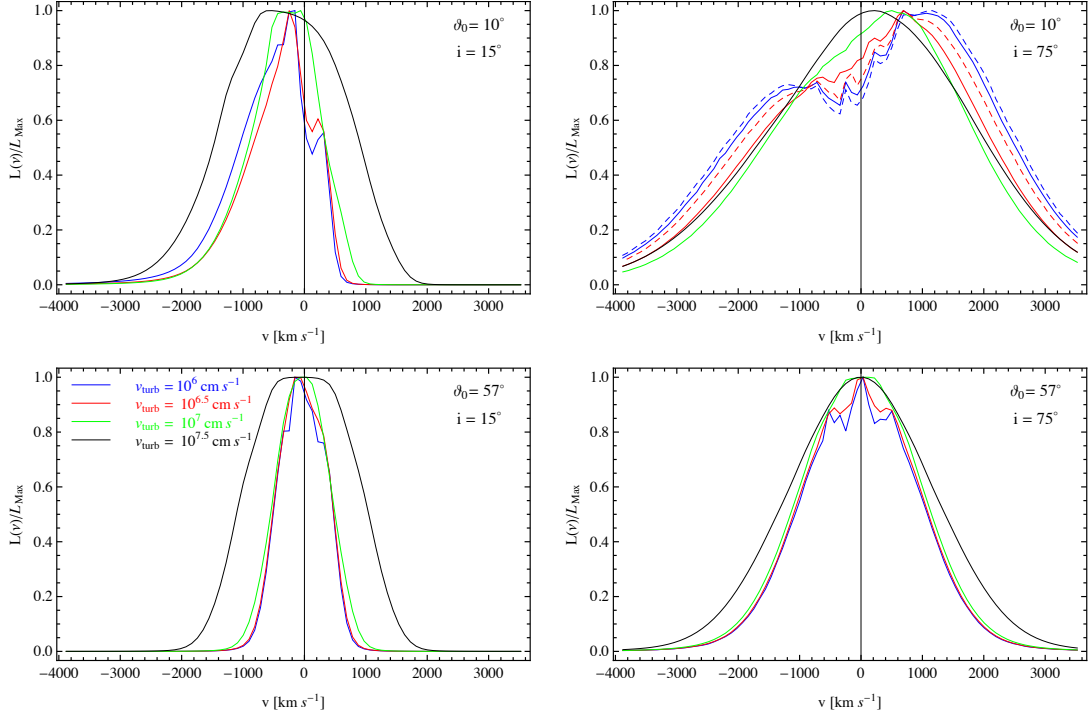} 
\caption{
$L_\nu/L_{\rmn {max}}$ versus velocity for 4 different values of the thermal plus turbulent velocity: 
$v_{\rmn {turb}} =10^6, 10^{6.5}, 10^{7}, 10^{7.5}$ cm s$^{-1}$ and two different inclination angles: 
$i= 15\degr$ (left panels) and $i=75\degr$ (right panels). Upper panels correspond to 
$\vartheta_0 = 10\degr$ and lower panels, to $\vartheta_0 = 57\degr$. The initial density is 
$n_0 = 10^{11}$ cm$^{-3}$ in all cases. The two extra curves shown in the upper right panel 
correspond to $v_{\rmn {turb}} =10^{5.8}$ cm s$^{-1}$ (dashed blue) and $10^{6.2}$ cm s$^{-1}$ 
(dashed red). 
}
\label{Fig:V_turb-var-new} 
\end{figure*} 

\begin{figure*}
\includegraphics[width=.825\textwidth,angle=0]{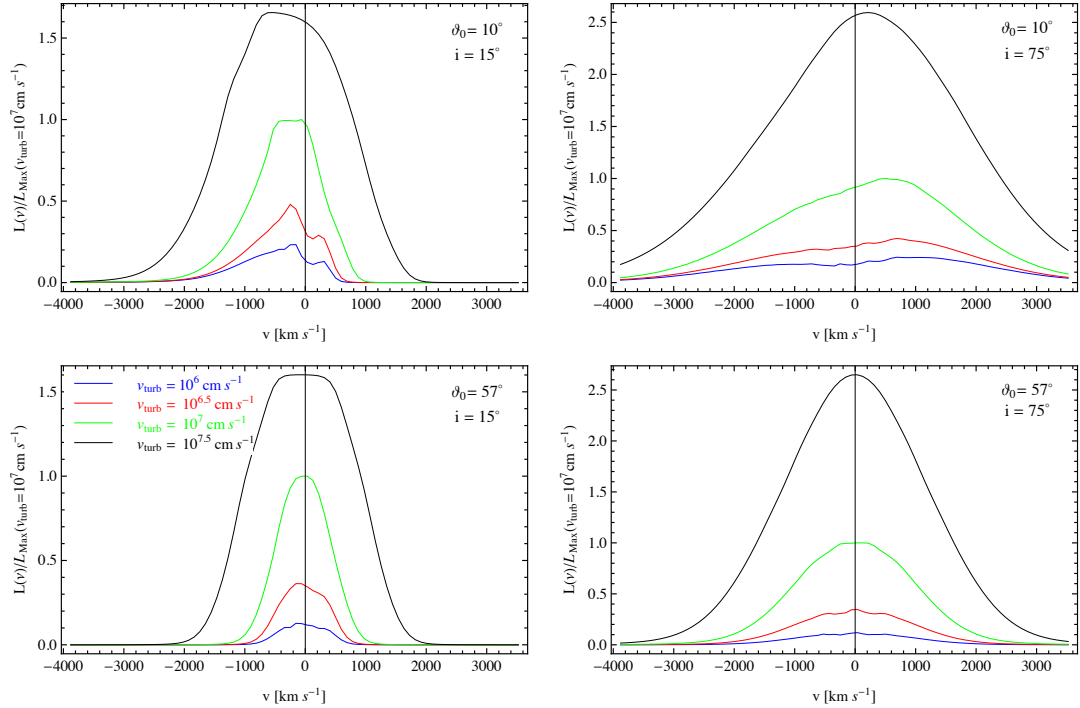}  
\caption{
$L_\nu/L_{\rmn {max}}(v_{\rmn {turb}}=10^7$ cm s$^{-1})$ versus velocity for  
four different values of the thermal plus turbulent velocity: 
$v_{\rmn {turb}} =10^6$, $10^{6.5}, 10^{7}, 10^{7.5}$ cm s$^{-1}$ and two different 
inclination angles: $i= 15\degr$ (left panels) and $i=75\degr$ (right panels). 
Upper panels correspond to $\vartheta_0 = 10\degr$ and lower panels, to 
$\vartheta_0 = 57\degr$. 
}
\label{Fig:V_turb-var-NormToVth70} 
\end{figure*} 

These results suggested considering different velocities by incorporating the effect of turbulence.
\citet{Bottorff&Ferland00} studied how microturbulence can affect the lines and 
showed that it affects more the far UV lines. 
Figure~\ref{Fig:V_turb-var-new} shows the lines corresponding to four different values of 
$v_{\rm turb}$, for two different inclination angles, $i=15\degr$ (left panels) and $i=75\degr$ 
(right panels). The profiles in the upper row correspond to $\vartheta_0=10\degr$ and in the lower 
row, to $\vartheta_0=57\degr$. In general, the lines become smoother and more symmetric  with 
increasing $v_{\rm turb}$.   
A much higher $v_{\rmn {turb}}$ does make a noticeable difference, as expected. However, 
note that the upper right panel seems to represent an anomalous situation, as the profiles 
become narrower as the turbulent velocity increases.  

To investigate the apparently anomalous situation in the upper right corner of 
Fig.~\ref{Fig:V_turb-var-NormToVth70} we plotted the same profiles as above, but 
normalized to the values corresponding to our fiducial turbulent velocity 
($10^7$ cm s$^{-1}$). Note that in all cases, even in the apparently deviant case, 
the flux increases when the turbulent velocity does.    

\subsection{Changing inclination angle at fixed launching angle} 
\label{sec: subsec5.2} 
\begin{figure*}
\includegraphics[width=0.925\textwidth,angle=0]{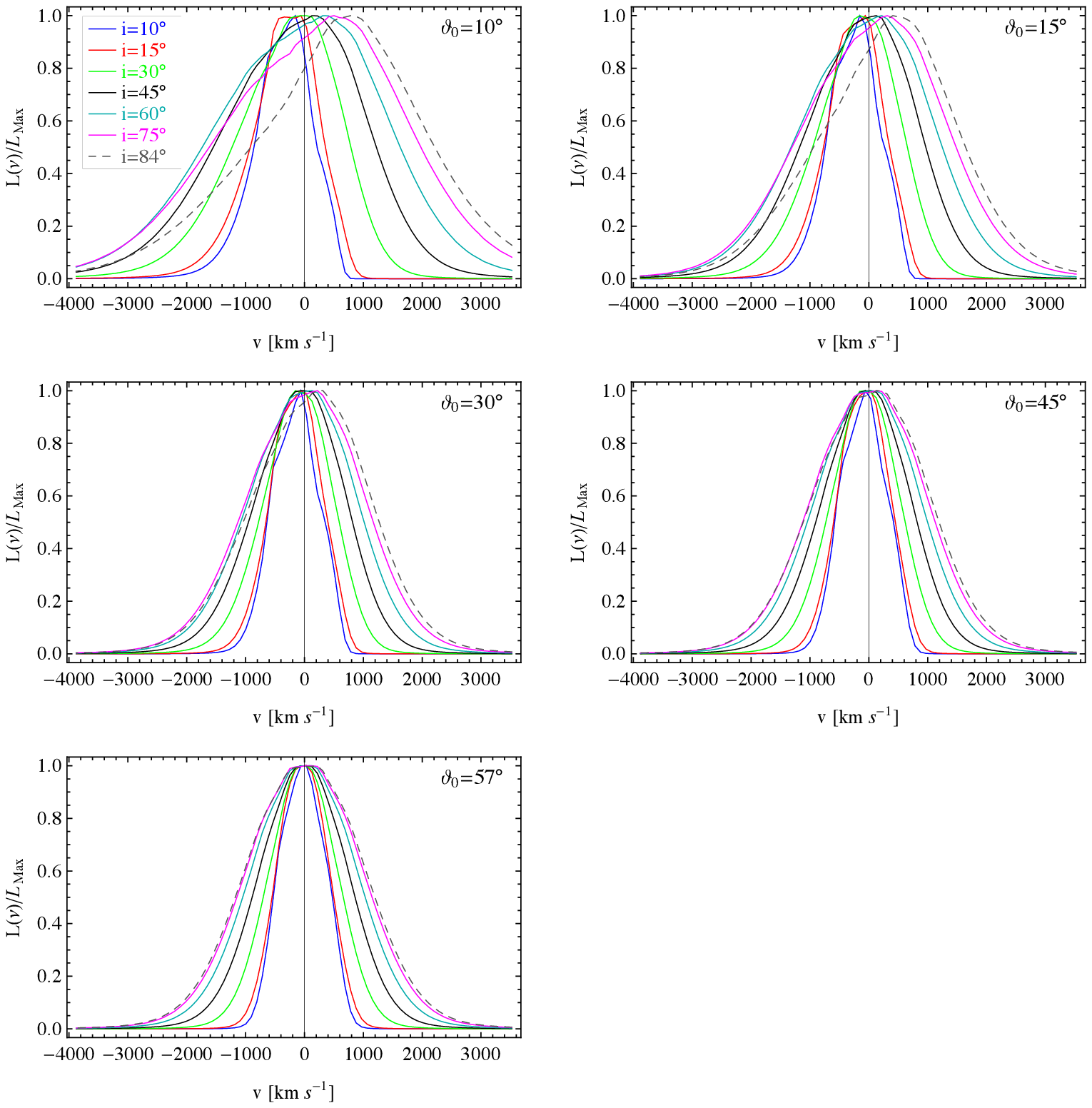} 
\caption{$L_\nu/L_{\rmn {max}}$ versus velocity for $i= 10\degr-84\degr$. Looking 
clockwise from the upper left: $\vartheta_0 =  10\degr, 15\degr, 30\degr, 45\degr, 57\degr $. 
The latter two cases do not differ significantly.
}
\label{FigSet:i25-to-78-Norm} 
\end{figure*}

For the fiducial values of density 
($n_0=10^{11}$ cm$^{-3}$) 
and turbulent velocity ($v_{\rm turb} =10^{7}$ cm s$^{-1}$),  we studied the effect(s) 
of changing the launching and viewing angles, in the ranges $\vartheta_0 = 5\degr-57\degr$ 
and $i = 10\degr-84\degr$, respectively. 
The results are shown in Figures  \ref{FigSet:i25-to-78-Norm} and \ref{Fig_Set:th15and45_i25to78}. 

In each panel of Figure \ref{FigSet:i25-to-78-Norm} the profiles are plotted 
versus velocity for a given launching angle, to enhance the effect of 
changing the viewing angle. 
Zero velocity is the average of the two doublet wavelengths.
The velocities plotted represent velocities from the observer's point of view,
therefore negative velocities correspond to blueshifts. 
In all cases the profiles are slightly asymmetric, with increasing degree of asymmetry 
with decreasing inclination. The blue wings change less than the red wings, so that 
as the inclination angle approaches to smaller values, the red wings are increasingly 
weaker. In Figure \ref{FigSet:i25-to-78-Norm}, the effect is hard to notice for the 
lowest $\vartheta_0$ (the two upper panels),  due to the shift to the red in the peak of 
the profiles when the inclination increases. 
We will discuss the effect of launching angle dependency below.

A way to see this is by noting that, for a given launching angle, when the object
is seen face-on, the projection of the velocity into the line of sight is towards the observer
for any azimuthal angle, while for objects seen edge-on, that projection is towards the
observer for part of the emitting region, and receding from them for the rest.  
For intermediate cases, the closer the object's LOS is to the 
face-on case, the more the red wing of its profile is weakened, explaining why the 
lines are less symmetric for smaller inclination angles.

\subsection{Changing launching angle at fixed inclination angle} 
\label{sec: subsec5.3} 

\begin{figure*}
\includegraphics[width=0.925\textwidth,angle=0]{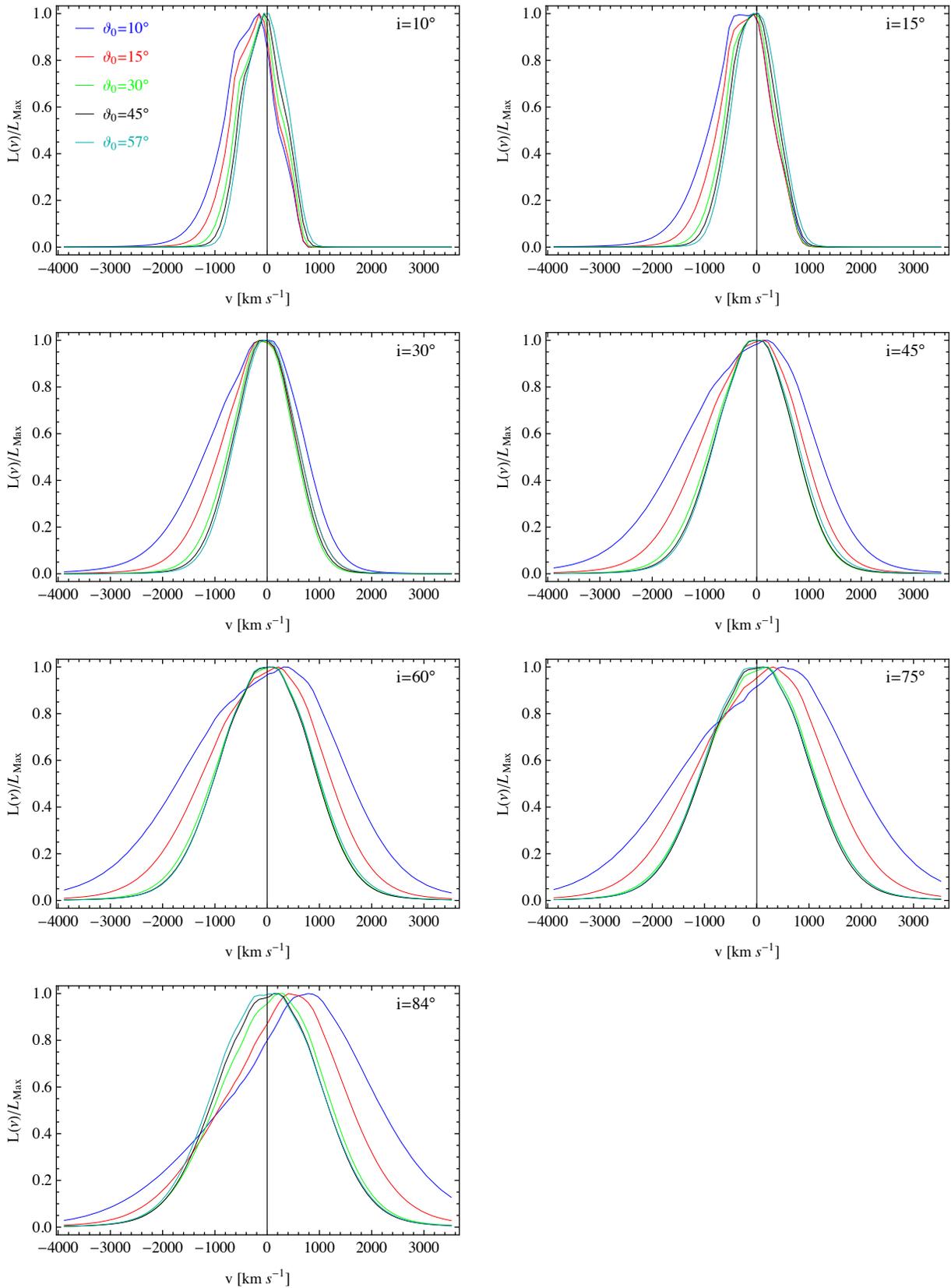} 
\caption{
$L_\nu/L_{\rmn {max}}$ versus velocity.  
In each panel, we have plotted the profiles corresponding to $i$ from $10\degr$ to
$84\degr$  for $\vartheta_0 = 10\degr- 57\degr$.  
For given $i$, those profiles for which $\vartheta > i$ are bluer, while those corresponding 
to $\vartheta < i$ are redder.  Here, $\vartheta > \vartheta_0$ is the angle between a line
launched with some $\vartheta_0$ and the base of the emitting region. 
}
\label{Fig_Set:th15and45_i25to78} 
\end{figure*}

The effects of changing the launching angle $\vartheta_0$ are shown in Figure
\ref{Fig_Set:th15and45_i25to78}, where in each panel we have plotted the
profiles for a given inclination angle and different launching angles.
The actual angle to be considered is the angle $\vartheta$ at which a line 
launched with some $\vartheta_0$ crosses the base of the emitting region
(when $\vartheta_0$ increases, so does $\vartheta$). For instance, 
for our chosen value of $\tan\beta$, 
for $\vartheta_0=20\degr$, $\vartheta \sim 25.82\degr$ and 
for $\vartheta_0=45\degr$, $\vartheta = 48\degr$

When  $\vartheta_0$ increases, the projection of the wind velocity onto the LOS is 
towards the observer in a portion of the emission region (i.e., for some azimuths)  
and is also towards the observer in the rest of the region as long as $\vartheta > i$.
In the cases $\vartheta < i$, that projection is receding from the observer.  
As the wind velocity decreases with increasing $\vartheta_0$, so does 
the magnitude of its projection for given $i$, and thus the blueshift decreases for 
increasing $\vartheta_0$. 
However, it is the Doppler velocity, including a contribution from the 
rotational velocity, which is the velocity relevant for producing the observed
line profiles. Thus, as the wind velocity decreases with $\vartheta$, then not 
only is the blueshift reduced, but the rotational velocity is increasingly 
dominant and the profiles become more symmetric. 
For any launch angle, the relative importance of the receding 
term with respect to the approaching term increases with increasing 
viewing angle, but the effect is larger for smaller $\vartheta_0$. 

We also analysed how strongly this broadening of line profile with decreasing 
$\vartheta_0$ depends on the density profile. Note that, in principle, the smaller 
$\vartheta_0$ is, the larger the radii at which the streamlines intersect the base 
of the emission region, that is, these lines will form at radii where the density 
(that goes as $\sim r^{-b}$) is smaller, affecting the optical depth.  
To that end, we compared the inverse square power-law to other (less steep) 
density power-laws ($b=0.5, 1$) for different launching angles, and found that 
the dependence is negligible. That is, the broadening found in the small 
$\vartheta_0$ cases depends mainly on the velocity projection. 

In summary, the relevant quantity is neither of the angles, but a combination of them.  
This can also be seen by considering the expression of the optical depth (Eq. \ref{eq: tau}) 
where the frequency dependance is encompassed in the exponent $x_\nu$, dependent
on the Doppler velocity $v_{\rmn {D}}$. The latter includes the wind contribution, which 
ranges from $v_{\rmn {p}} \sin(\vartheta + i)$ when $\phi=0$, to $v_{\rmn {p}} \sin(\vartheta-i)$ 
when $\phi=\pi$.  
Note also, that the FWHM increases with increasing inclination angle (see 
Fig.~\ref{FigSet:i25-to-78-Norm}), but for small launching angles it reaches 
its maximum at $i < 84\degr$, while the maximum is reached at $i=84\degr$ 
for larger $\vartheta_0$.  In the smaller $\vartheta_0$ cases, the broadening 
and subsequent decrement is accompanied by a shift in the peak, from bluer 
(at smaller inclinations) to redder velocities (at larger inclinations). This is due 
to the fact that the observer sees the base of the conical emission region from 
an almost edge-on perspective, so that the part of the cone with $\phi\simeq 0\degr$ 
(which produces blueshifted emission) has very small projected surface area.

\subsection{Changing density}
\label{sec: subsec5.4} 

We also looked at the effects of varying the initial density. 
Although we adopted $n_0 \sim 10^{11}$ cm$^{-3}$ as the ``standard density'', we 
also chose to check the effect of even lower and higher densities. In principle, one 
would expect broader profiles for smaller initial density. In fact, that is what is found 
when running simulations that do not include the terms and factors introduced in Hall 
et al. (2012). In that case, the results  showed that the profiles become broader as the 
initial density decreases. 
In effect, as the density decreases, so does the opacity and, in that case there will be 
less photons absorbed in the line wings and  this translates into broader lines. 
However, the inclusion of these previously neglected terms and factors modifies  
the behaviour of the profiles, in such a way that the effect of changing the initial 
density is much less important (negligible, in some cases). In the current model, the 
velocity field (which depends on both inclination and launching angles) dictates the 
optical depth behaviour. This is somewhat similar to the broadening of the 
low-$\vartheta_0$ case that we discussed above.  

\begin{figure*} 
\includegraphics[width=.750\textwidth,angle=0]{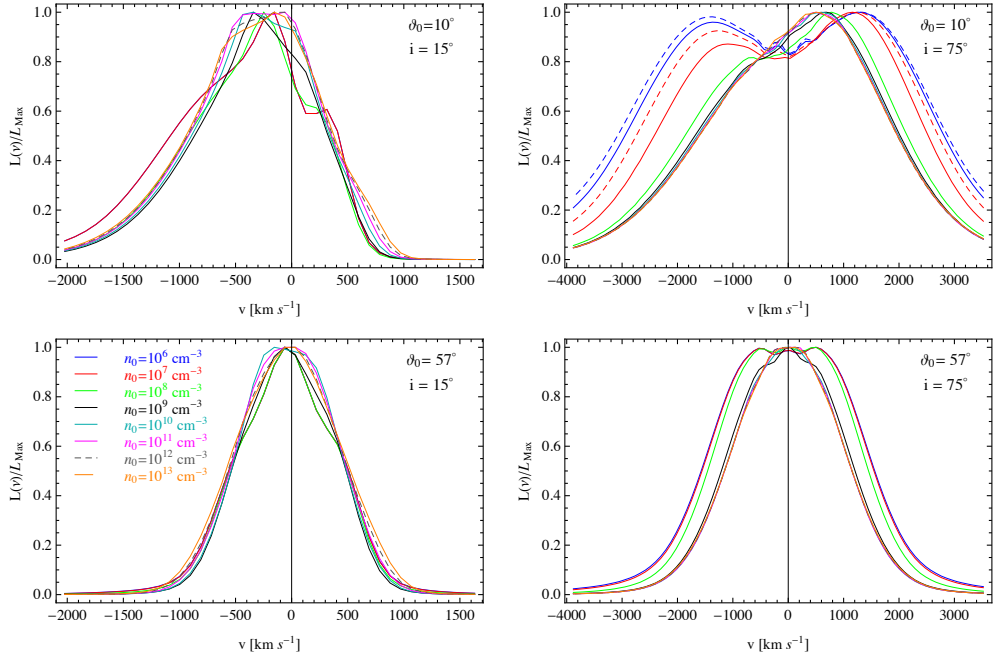} 
\caption{
$L_\nu/L_{\rmn {max}}$ versus velocity. Here, the profiles correspond to fixed 
$i\sim 15\degr$ (left panels) and $i\sim 75\degr$ (right panels) and 
$\vartheta_0 = 10\degr$ (upper panels) and $\vartheta_0 = 57\degr$ 
(bottom panels) but different initial densities: $n_0 = 10^6$ cm$^{-3}$  (blue),
$n_0 = 10^7$ cm$^{-3}$ (red), $n_0 = 10^8$ cm$^{-3}$ (green), 
$n_0 = 10^{9}$ cm$^{-3}$ (black), $n_0 = 10^{10}$ cm$^{-3}$ (cyan), 
$n_0 = 10^{11}$ cm$^{-3}$ (magenta), $n_0 = 10^{12}$ cm$^{-3}$ (dashed gray) and 
$n_0 = 10^{13}$ cm$^{-3}$ (orange). The extra lines in the upper right panel correspond 
to $n_0 = 10^5$ cm$^{-3}$  (dashed blue) and $n_0 = 10^{6.5}$ cm$^{-3}$  (dashed 
red) and were included to show that, although slower than in other cases, the low-density 
profiles also converge.  
The effect of a lower density on the opacity, and thus on the line broadness, is surpassed 
by the effect of the velocity field, leaving a weak dependence on density, especially at lower 
inclinations. For fixed $\vartheta_0$ (i.e., looking along rows) the spread is larger for 
higher inclination angles, while for fixed $i$ (i.e, looking along columns), it is larger for 
smaller launching angle, with the trend being more pronounced with decreasing $n_0$. 
}
\label{fig: Set:n0var-11Aug02} 
\end{figure*}

Figure~\ref{fig: Set:n0var-11Aug02} shows the profiles obtained for a fixed inclination 
$i = 15\degr$ (left panels) and $i = 75\degr$ (right panels) and launching angles 
$\vartheta_0 = 15\degr$ (upper panels) and $\vartheta_0 = 57\degr$ (lower panels) 
when the initial  density, which declines radially according to $n \sim r^{-2}$ is 
varied. 
\footnote{Note that the velocity ranges are not the same in all cases:  the left panels 
share the velocity range, but that differs from the range in either of the two right panels.} 
The results of the density analysis also show that for given $\vartheta_0$, the smaller 
the inclination angle, the bluer the maximum.   
Note also that we have included two extra profiles in the upper right panel, 
to illustrate that in this particular case ($\vartheta_0 = 10\degr$,  $i = 75\degr$) the profiles 
do indeed converge at lower densities. 

\subsection{Smaller launching angles} 
\label{sec: subsec5.5} 

\begin{figure*} 
\includegraphics[width=.750\textwidth,angle=0]{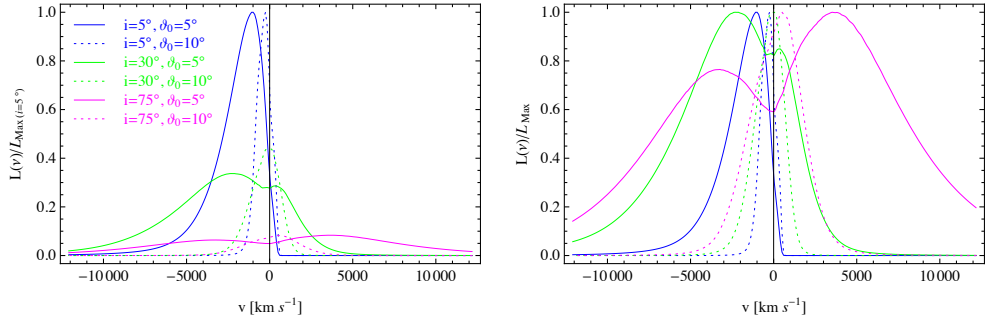} 
\caption{Two different normalizations of the profiles for  $\vartheta_0 = 5\degr, 10\degr$ 
and $i= 5\degr, 30\degr, 75\degr$. In the left panel, the profiles are normalized to the 
maximum of the $i=5\degr$ for the corresponding launching angle. In the right panel, 
the normalization is with respect to each profile's own maximum. 
}
\label{Fig_Set:th0small} 
\end{figure*}

The effect of even smaller launching angles is shown in Figure \ref{Fig_Set:th0small}
for the cases $i= 5\degr, 30\degr, 75\degr$. Included, for comparison, are 
the profiles for the same inclination angles, but with $\vartheta_0=10\degr$. 
In the left panel, each profile is normalized with respect to the maximum of the 
$i=5\degr$ profile for each launching angle, whereas in the right panel the 
normalization is with respect to its own maximum. 
Two observations can be made from the figure. First,  the differences between 
$\vartheta_0=5\degr$ and $\vartheta_0=10\degr$ profiles are much larger than 
those between $\vartheta_0=10\degr$ and $\vartheta_0=15\degr$ profiles.  
Second, when the viewing angle is large and $\vartheta_0 = 5\degr$, the profile 
is double-peaked, which is not observed in \ionCiv \, lines.  

\begin{figure*} 
\includegraphics[width=.750\textwidth,angle=0]{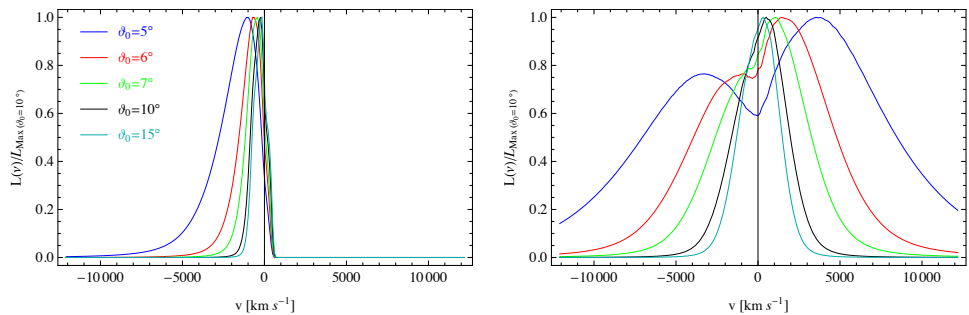} 
\caption{Normalized profiles for  $\vartheta_0 = 5\degr$-$7\degr, 10\degr,15\degr$ 
and $i= 5\degr$ (left panel), $75\degr$ (right panel). 
}
\label{Fig_Set:th0small-new} 
\end{figure*}

To study these issues we analysed the evolution of the profiles, for two 
different inclination angles ($i=5\degr, 75\degr$), when the launching 
angle changes between $\vartheta_0=5\degr$ and $15\degr$. 
Figure \ref{Fig_Set:th0small-new} shows that in both cases, as the launching 
angle increases the profiles become increasingly narrower. 
Here we see the same trend shown in Figure \ref{Fig_Set:th15and45_i25to78} and 
discussed in subsection \ref{sec: subsec5.3}: for smaller launch and viewing 
angles (left panel), most of the flux is due to motion towards the observer and
the blueshift decreases with increasing $\vartheta_0$.  For larger $i$ (right
panel) the contribution of the receding term of the Doppler velocity dominates
but also decreases with increasing $\vartheta_0$, leading to an emission peak
that approaches the systemic redshift with increasing $\vartheta_0$.  
Also noticeable is the fact that the double-peaked feature is only present when 
$\vartheta_0=5\degr$ (and to lower extent when $\vartheta_0=6\degr$). 
That suggests that we can impose an empirical restriction on $\vartheta_0$ and 
consider only those that satisfy the condition $\vartheta_0 > 6\degr$. 
To determine whether this is a general constraint, valid for any given $\beta$,
would require simulations for different values of that parameter,
which is beyond the scope of this work.    

\section{Line Width Measures} 
\label{sec:FWHMs} 

From a set of profiles obtained for different inclination angles $i$ 
and launching angles $\vartheta_0$ we study how the FWHMs are 
distributed as a function of the angles $i$ at which quasars are visible. 
To do so, we use an approach similar to that of \citet{Fine+08},
who constrained the range of possible AGN viewing angles by using 
geometrical models for the BLR and comparing the expected 
dispersion in linewidths at each viewing angle to their observational data. 
We extend their analysis by also considering the clumpy torus model 
of Nenkova et al. (2008a, 2008b; hereafter N08), 
as constrained by \citet{Mor+09} using infrared observations of luminous AGN.

\citet{Fine+08} measured the linewidth of the \ionMgii \, line in 32214 quasar 
spectra from the Sloan Digital Sky Survey (SDSS) Data Release Five, 2dF QSO
Redshift survey (2QZ) and 2dF SDSS LRG and QSO (2SLAQ) survey 
and found that the dispersion in linewidths strongly correlates with the 
optical luminosity of QSOs. 
\citet{Fine+10} used 13776 quasars from the same surveys to study the dispersion
in the distribution of \ionCiv \, linewidths. In contrast to their findings for
the \ionMgii, they found that the dispersion in \ionCiv \, linewidths is
essentially independent of both redshift and luminosity. 
\citet{Fine+08, Fine+10} used the fact that if the linewidth measured from a 
spectrum depends on the viewing angle to the object, 
the linewidth dispersion for a model for the BLR can by calculated 
by `observing' that model over different ranges of viewing angles.
Combinations of models and viewing angle ranges that 
give dispersions larger than the observed dispersion it can be rejected. 
Fine et al. assumed a coplanar obscuring torus surrounding the central SMBH 
and BLR with an opening angle $i_{\rmn {max}}$ (measured from the vertical 
axis), so that the viewing angle $i$ should satisfy $i \leq i_{\rmn {max}}$. 
If the FWHM of a BEL varies with $i$, the dispersion in the FWHM distribution 
of that BEL should vary with $i_{\rmn {max}}$. 

\subsection{Fine et al. test}
Following Fine et al. (2008), we compare 
the dispersion of observed $\log(\text{FWHM})$ values 
with the dispersion of our simulated $\log(\text{FWHM})$ 
as a function of $i_{\rmn {max}}$ and launching angle $\vartheta_0$ 
to see if we can constrain $i_{\rm max}$ or $\vartheta_0$. 

Using the launching angle as a parameter, we evaluate the dispersion 
of the function $f(i)=\log(\text{FWHM}(i))$. 
As all the variables are, in fact, continuous, we interpolated each set 
of FWHMs to obtain the corresponding  continuous functions. 
For \textit {a given} $i_{\rmn {min}}$, the mean and the variance of the 
FWHMs are functions of $i_{\rm max}$, according to 
\begin{align}
\bar{f}(i_{\rmn {max}})  = &
\frac{ \int_{i_{\rmn {min}}}^{i_{\rmn {max}} }  \sin i \; P(i)\; f(i) \, di}
{ \int_{i_{\rmn {min}}}^{i_{\rmn {max}} }  \sin i \; P(i) \, di}  
\label{eq: mean-f} \,, \\
\sigma^2_f(i_{\rmn {max}})  = &   
\frac{ \int_{i_{\rmn {min}}}^{i_{\rmn {max}} }  \sin i \; P(i) \;  \left[ f(i)-\bar{f}(i_{\rmn {max}}) \right ]^2 \, di}
{ \int_{i_{\rmn {min}}}^{i_{\rmn {max}} }  \sin i \; P(i)\, di} \, ,  
\label{eq:disp-f}
\end{align} 
where 
$P(i)$ is a weighting factor, equal to 1 in the \citet{Fine+08} approach, 
that measures the probability of not having obscuration in the LOS direction.  
We first present results using $P(i)=1$ and then turn to a more complex case. 
Both \citet{Fine+08} and \citet{Mor+09} have  $i_{\rmn {max}}=90\degr$ 
as the upper limit for that angle. However, we have an extra limitation, 
set by the inclination of the base of the emitting region, chosen to 
be $\beta=6\degr$. Therefore, our upper limit is $i_{\rmn {max}}=84\degr$.      

Noting that  \citet{Fine+08, Fine+10} have employed inter-percentile values (IPVs) 
rather than FWHMs to characterize the line widths, we also investigated the 
behaviour of this line measure from our results. 
For a given percentage $p$, the definition of IPV$p$ suggested by \citet{Whittle85a} 
is the separation between the median (where the integrated profile reaches the 
$50\%$ of the total flux) and the positions where $p\%$ and $(100-p)\%$ of the 
total flux are reached. Thus, calling $d_1$ and $d_2$ the distances between the 
median and $p$ and $(100-p)$ respectively, IPV$p = d_1+d_2$. 
Note that EBS92 define the analogous quantity $W_x$ (half-width at $x$), where 
$x$ is defined as a given fraction of the peak flux. 

\begin{figure*} 
\includegraphics[width=0.725\textwidth,angle=0]{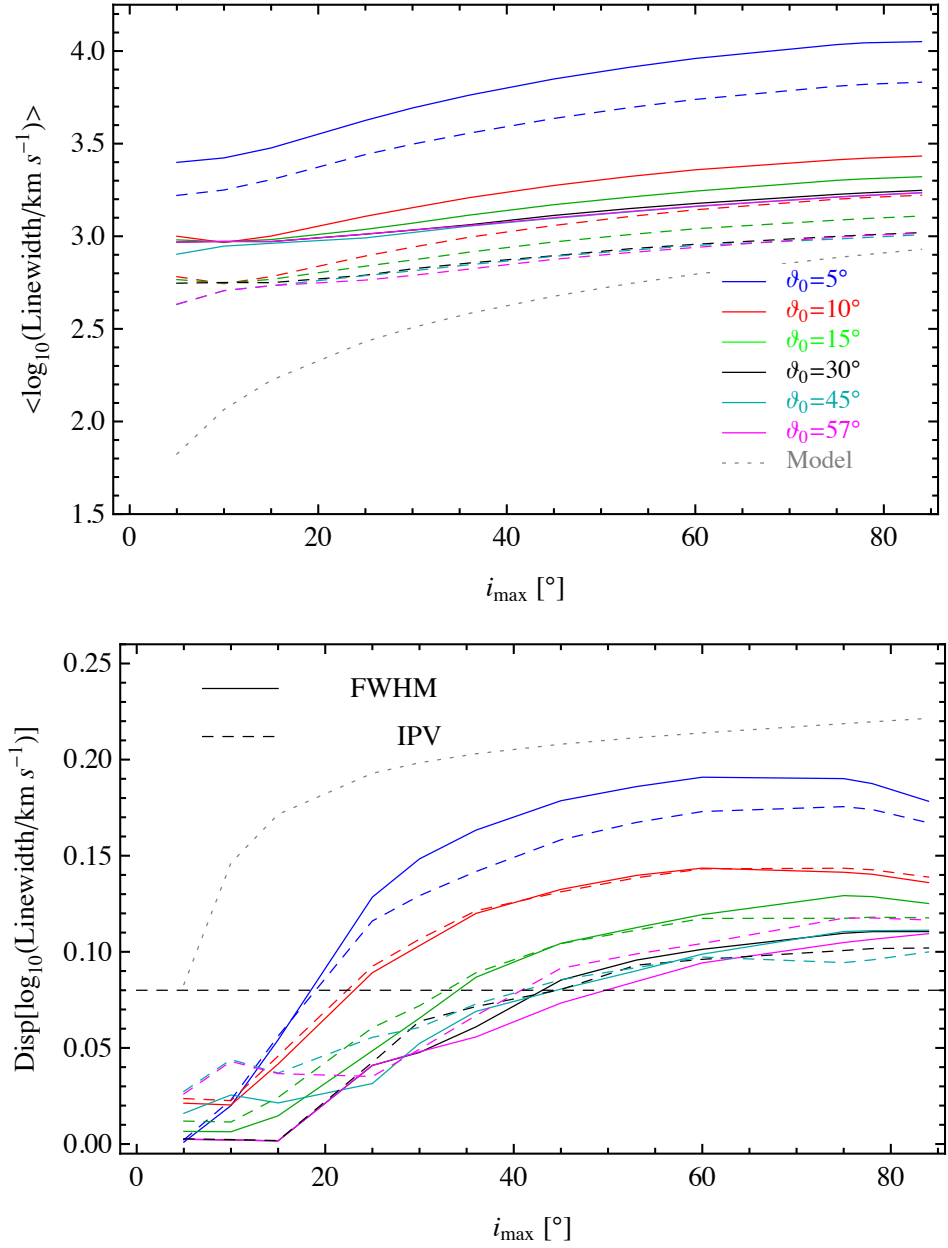} 
\caption{
Averages (top panesl) and dispersions (bottom panel) of $\log(\text{FWHM})$ (solid lines) 
and $\log(\text{IPV25})$ (dashed lines) evaluated for different launching angles, using 
$P(i)=1$. The minimum viewing angle is $i_{\rmn {min}}=2.5^{\circ}$. Included is the 
plot (dotted curve) of the dispersion for a model of the form 
$g(i) = \text{FWHM}(i) = i(1000 \text{ km s}^{-1})$ (see text).
The dashed horizontal line plotted together with the dispersions corresponds to the 
\citet{Fine+10} results,  its  meaning is commented below. 
}
\label{Fig:Disp-FWHMandIPV-12May16} 
\end{figure*}

Figure \ref{Fig:Disp-FWHMandIPV-12May16} shows the averages (top panel) 
and dispersions (bottom panel) obtained for  $\vartheta_0 = 5\degr, 10\degr, 
15\degr, 30\degr, 45\degr, 57\degr$ for different  $i= i_{\rmn {max}}$ ranging 
from $5\degr$ to $84\degr$ when $i_{\rmn {min}} = 2.5\degr$. Solid lines 
correspond to FWHM and dashed lines to IPV line width measurements. 
For most maximum inclination angles, for fixed $i_{\rmn {max}}$, the dispersion 
of the FWHMs decreases with increasing $\vartheta_0$. In particular, the 
dispersion of FWHM for the smallest launching angle is systematically larger 
than for all others. 
We can see that the general trend is lower dispersion for higher $\vartheta_0$, 
except for the smaller $i_{\rmn {max}}$, where  
$\sigma_{\vartheta_0 = 45\degr}$ departs from it. 
\footnote{We denote 
$\sigma_{\text{FWHM}(\vartheta_0 = x\degr)}$ by $\sigma_{\vartheta_0 = x\degr}$.} 

We also compared these results to a simple model which assumes
$g(i)= \text{FWHM}(i) = i (\,1000$ km s$^{-1}$), merely to see how the dispersion
in log(FWHM) behaves for a model with known variation of FWHM with $i$.
The dotted line in Figure \ref{Fig:Disp-FWHMandIPV-12May16} corresponds to this
simple model. For the $i_{\rmn {min}}$ considered in the figure, this model departs 
from the observational results at any $\vartheta_0$. 
We found that when $i_{\rm min} = 7.5\degr$, the model approximately matches 
the result for $\vartheta_0=5\degr$ in the range $ 35\degr \lesssim i_{\rmn {max}} 
\lesssim 60\degr$. 
In the general case, it can be inferred that a more sophisticated model is needed. 
Such a model probably has to include information about the launching angle.

The dispersions obtained for the FWHMs are increasing functions of the 
parameter $i_{\rmn {max}}$, although they show a mild decreasing trend 
at $i_{\rmn {max}} \gtrsim 60\degr$. Similarly, the dispersions of the IPVs are increasing 
functions of  $i_{\rm max}$. Also included in 
Figure \ref{Fig:Disp-FWHMandIPV-12May16} is the 0.08 dex dispersion line reported by
\citet{Fine+10} as the observational upper limit on the dispersion measured from their sample.
We can see that as the torus half-opening angle (measured from the polar axis, 
and represented by $i_{\rmn {max}}$) increases above about $18\degr$, the 
wind launch angles required to match the \citet{Fine+10} constraints are 
increasingly larger. 

\begin{figure*} 
\includegraphics[width=.850\textwidth,angle=0]{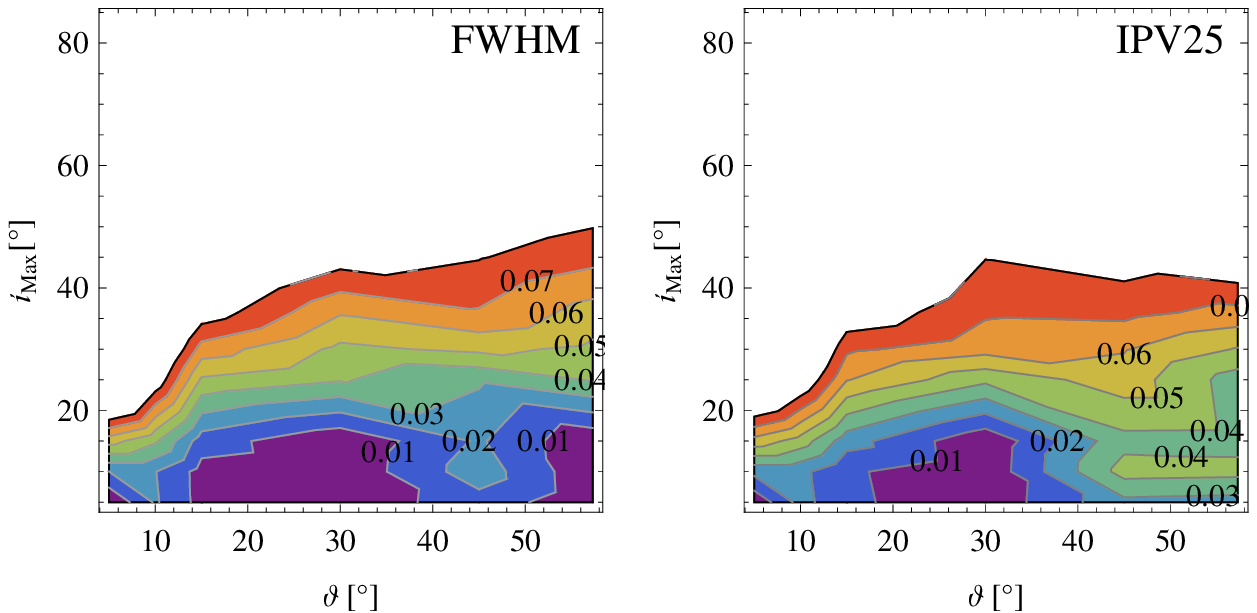} 
\caption{
Contour plot of the standard deviation of the FWHM vs launching and 
inclination angles. Only the contours within the region that 
matches the Fine et al. results are shown.  
}
\label{Fig: ContourPlot-FWHMandIPV25-11Sept22} 
\end{figure*}

Figure \ref{Fig: ContourPlot-FWHMandIPV25-11Sept22} shows the allowed 
region in the $i$-$\vartheta_0$ plane when analysing the dispersion of FWHMs 
(left panel) and of IPV25s (right panel). In both cases, $i_{\rmn {min}} = 2.5\degr$.   
Our results give, within the  $\vartheta_0<60\degr$ range 
allowed by the MHD solutions, a maximum half-opening angle of about 
$47\degr$, above which no wind launch angle matches the observations. 
This maximum torus half-opening angle has a somewhat different behaviour 
if the IPVs are considered, reaching a maximum at $\vartheta_0 \sim 30\degr$ 
and declining for larger $\vartheta_0$. However, note that the ``absolute'' 
maximum is similar in both cases. 

In section \ref{sect: Res} we showed that the profiles obtained for cases 
corresponding to larger inclination and small launching angle had 
double-horned profiles and mentioned that this contradicts observational 
results. The analysis presented in this section shows that such combinations 
are in fact ruled out. 

\subsection{Clumpy torus} 

\begin{figure*} 
\includegraphics[width=.850\textwidth,angle=0]{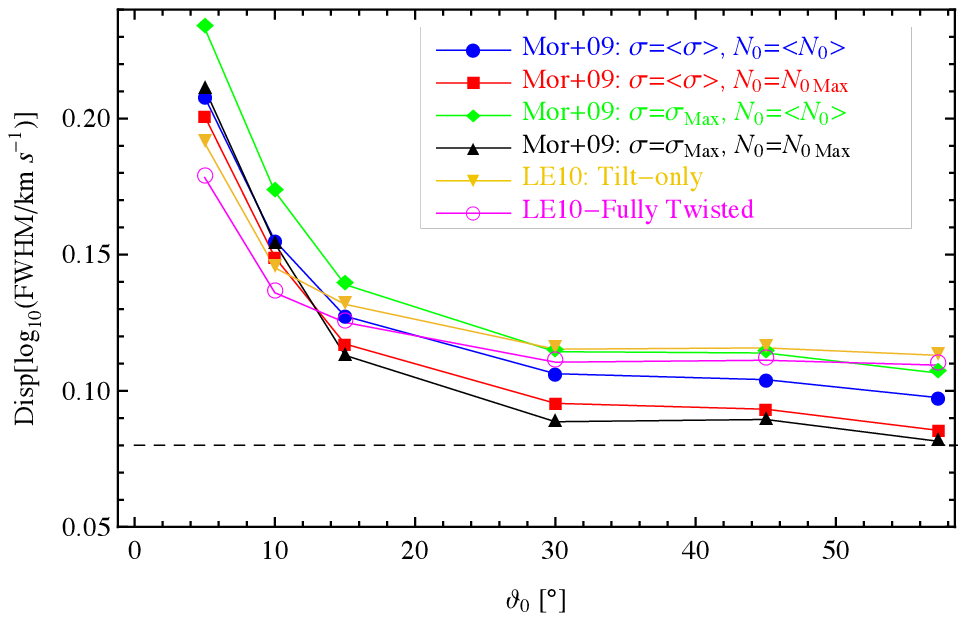}  
\caption{
Dispersions of  $\log(\text{FWHM})$ using $P(i= \pi/2-\beta)$ as given by 
Eq. \ref{eq:Pesc}. Here, $i_{\rmn {min}} = 2.5\degr$, $i_{\rmn {max}} = 84\degr$.  
Each of the lines obtained using the \citet{Mor+09} prescription represents 
a different combination of $\sigma$ and $N_0$. Also included is a line 
corresponding to dispersions calculated assuming a tilt-only warped disc 
(see next section). The dashed line corresponds to the \citet{Fine+10} results.  
}
\label{Fig:DispersionsFWHM-Mor09} 
\end{figure*}

As mentioned above, \citet{Mor+09} adopted the more detailed expression 
for the escape probability proposed by N08.    
In that model, the torus is clumpy, consisting of optically thick clouds and 
the quasar is obscured when one of such clouds is seen along the LOS. 
The torus is characterized by the inner radius of the cloud distribution 
(set to the dust sublimation radius,  $R_{\rmn {d}}$, that depends on the grain 
properties and mixture) and six other parameters. 
Equation (3) in \citet{Mor+09} provides the weighting factor that we used in 
our evaluation: 
\begin{equation} 
 P_{\rmn {esc}}(i) = \exp\left[{-N_{0}\exp\left({-\frac{(\pi/2-i)^{2}}{\sigma^{2}}}\right)}\right]
 \label{eq:Pesc}
 \end{equation} 
where $N_0$ is the mean number of clouds along a radial equatorial line 
and $\sigma$ is the torus width parameter (analogous to its opening angle).  
Implicitly, it is assumed that the disc and the torus are aligned. In their Fig. 6, 
\citet{Mor+09} present the torus parameter distributions for their sample, 
and from there it is clear that the distribution of the two parameters we need 
to input in Eq. \ref{eq:Pesc} (namely, $N_0$ and $\sigma$) are very broad. 
For completeness, we have reproduced in Table \ref{Tab: Table2} the minimum, 
mean and maximum values of the two parameters, taken from \citet{Mor+09}. 
Note that within this model, we only have $i_{\rmn {max}}=\pi/2$. The resulting 
distribution of the dispersions with the launching angle are presented in Figure 
\ref{Fig:DispersionsFWHM-Mor09}, where each line corresponds to a given 
combination of $\sigma$ and $N_0$. For clarity, in the figure we excluded 
combinations such that at least one of the parameters takes its minimum value, 
as such combinations yield lines farther away from the observed upper limit dispersion.

\begin{table*} 
\caption{Minimum, mean and maximum values of the $N_0$ and $\sigma$ 
torus parameters from the \citet{Mor+09} sample.}
\label{Tab: Table2}
\begin{tabular}{@{}cccc} 
\hline
Parameter &   Minimum  &  Mean  &  Maximum \\
\hline
$N_0$ & 1 & 4.923 & 8 \\
$\sigma$ [$\degr$]  &  15  & 34 & 57 \\ 
\hline
\end{tabular}
\end{table*}%

For $\vartheta_0 \lesssim 30\degr$, all cases depart significantly from the Fine at al. 
result. For $\vartheta_0 \gtrsim 30\degr$, most cases remain highly incompatible 
with the Fine et al. results, but the curves are closer to them when 
$N_0=\max(N_0^{\rmn {Mor+09}})$, with $\sigma=\max(\sigma^{\rmn {Mor+09}})$ 
and $\sigma=\bar{\sigma}^{\rmn {Mor+09}}$.    
That motivated us to look for combinations of these parameters such that the results 
based on \citet{Mor+09} model match the Fine et al. limit, at least for some launch 
angles. Adopting $\sigma=\bar{ \sigma}^{\rmn {Mor+09}} $ and increasing $N_0$, 
one finds that for $N_0 \sim 15$ (the likely uppr limit, according to N08), all dispersions 
corresponding to 
$\vartheta_0 \gtrsim 30\degr$ are below the Fine et al. boundary and the rest have 
decreased in a similar amount (the effect is that the curve has almost rigidly moved 
down). 
When $N_0\sim 30$, only dispersions corresponding to $\vartheta_0 \lesssim 15\degr$ 
are above the Fine et al. boundary.  If, instead, $\sigma=\max(\sigma^{\rm Mor+09})$ 
is adopted, for $N_0 \sim 8.5$ the dispersion corresponding to 
$\vartheta_0 \sim 57\degr$ is already below the Fine et al. limit, and for 
$N_0 \sim 13.5$, all dispersions for the cases $\vartheta_0 \gtrsim 15\degr$ are 
below that line. 
On the other hand, keeping $N_0=\max(N_0^{\rm Mor+09})$ and increasing 
$\sigma$ does not lead to an improvement. 

Note that \citet{Mor+09} found $P(i=50\degr)\simeq 30\%$ and 
$P(i=70\degr)< 3\%$ when $N_0$ and $\sigma$ were set to their 
mean values, and, based on that, suggested that the inclination angle 
for type-1 objects should lie in the range $0\degr-60\degr$. 
However, the authors found that for the case in which all the parameters 
but the torus width are set to their mean values, the escape probability 
falls rapidly if $\sigma > 45\degr$.  
Our results indicate that the parameter $N_0$ is important when considering 
the dispersions of the line widths. In effect, as mentioned above, the only 
curves that are relatively close to the \citet{Fine+10} constraints correspond 
to the case $N_0=\max{(N_0^{\rm Mor+09})}$ and to have a better match 
larger $N_0$ values are needed. In that case, the $\sigma$ values should 
still be close to or larger than the mean from the \citet{Mor+09} sample. 

These results were obtained from a set of profiles corresponding to both mass 
and luminosity fixed, whereas  \citet{Fine+10} and \citet{Mor+09} samples 
involve a range of masses and luminosities. However, as already mentioned, 
\citet{Fine+10}  found that the dispersion in \ionCiv \, linewidths essentially does 
not depend on luminosity. As can be seen from their Figure 2, the IPV linewidth 
measurements are bound by 
$10^{8.25} \lesssim M_{\rmn {BH}}/M_{\sun} \lesssim 10^{10}$ 
and $0.1 \lesssim L/L_{\rmn {Edd}}  \lesssim 1$. 
Based ion that, we then anticipate that our results would not be strongly 
affected by considering different masses and/or luminosities.   

\section{Warped Discs}
\label{sect:warped discs}

As mentioned in the Introduction, the model of \citet[][LE10 hereafter]{LE10} 
replaces the torus by a warped disc.  In this section we explore whether we 
can infer new constraints on the BLR or on the parameters of warped discs 
by applying a restricted set of such warped disc models. 
Briefly, we evaluate the unobscured solid angle distribution as a function of 
observer inclination $i$, $dC(i)$, calculated for arbitrary disc tilt angle $\theta$. 
Then, we restrict our attention to the subset within our constraint $i < \pi/2 - \beta$, 
using the calculated unobscured solid angle distribution to determine the probability 
of the object being unobscured. Finally, we apply that probability to our emission 
line profiles, in a way analogous to that employed with the \citet{Fine+10} model. 

LE10 studied the fraction of type 2 AGN, $f_2 \approx 0.58$ among all AGN, and 
proposed a framework to account for it. 
They assumed that randomly directed infalling material at large scales would 
produce a warped disc at smaller scales, where it eventually aligns with the 
inner disc. They analysed both fully twisted and tilt-only cases (explained in 
detail below) under that assumption and showed that fully twisted discs can 
not reproduce the observed $f_2$. 
Models assuming tilt-only discs, on the other hand, match the observed $f_2$. 

A warped disc can be analysed as a series of annuli, each characterized by its 
radius and the two angles $\theta(r)$ (the angle between the spin axes of the 
annulus and the inner planar disc, i.e., the tilt) and $\phi(r)$ (the angle of the line 
of nodes measured with respect to a fixed axis on the equatorial plane, i.e., the 
twist). Thus, a fully twisted disc corresponds to $\phi(r) = 0$, $\phi(r + \delta r) = 2 \pi$, 
$\delta r \ll r$ and a tilt-only disc corresponds to the case of constant $\phi(r)$.
Each of the two warp modes can be associated to a covering factor $C$ depending 
on the misalignment, and the distribution of covering factors can then be inferred from 
the probability distribution of  the misalignment. Conversely, knowing the distribution 
of covering factors, the probability distribution of misalignments can be evaluated. This 
is the approach we take below. 
Note that LE10 calculated the azimuthally integrated covering factor while we study 
the covering fraction as a function of the azimuth. 

Consider first the tilt-only case. In Appendix  \ref{app-sect:unobsc} we derive the 
expression for the differential covering factor $dC(i)$ corresponding to this case. 
We estimated the unobscured fraction $P(i)=dC(i)/2\pi\sin(i)$ as a function of the 
tilt angle and found that considering a random distribution in solid angle of $\theta$ 
up to $\theta_{\rmn {max}} = \pi$ yields results incompatible with our line profiles. 
This is because that case corresponds to $i_{\rm max} = \pi/2$ for the Fine et al. 
case in all our diagrams.  
In a random distribution of incoming orientations with $\theta_{\rmn {max}} = \pi$, for 
every case of orientation $\theta$ there is a corresponding case with orientation 
$\pi - \theta$ (statistically speaking), which means that this model is identical to the 
case $i_max=pi/2$ for our purposes. 
A variant of that model, with a random distribution in solid angle of $\theta$ up to 
$\theta_{\rmn {max}} = \pi/2$ was also analysed.  
Using equation \ref{eq:Prob-Unobsc--LE10-tilt-only-ThetaMax90} for the probability 
and equations \ref{eq: mean-f} and \ref{eq:disp-f} we performed the same analysis 
applied in the Fine at al.  case.  
Note that the only angle to be considered in this case is $i_{\rm max}= \pi/2$, so 
we also ruled out this model.  %
We have included in Figure \ref{Fig:DispersionsFWHM-Mor09} the resulting 
dispersions for this model. The dispersions are far from the \citet{Fine+10} 
observational results for any launch angle considered. 

An analogous analysis was performed for the full twist case, with 
$0\leqslant \theta \leqslant \pi/2$. The probability of being unobscured for 
random inclinations is given by $P(i) = \cos(i)$ and, again, the region in 
the $i-\vartheta_0$ plane is the same obtained using other prescriptions. 

\begin{figure*}
\includegraphics[width=.850\textwidth,angle=0]{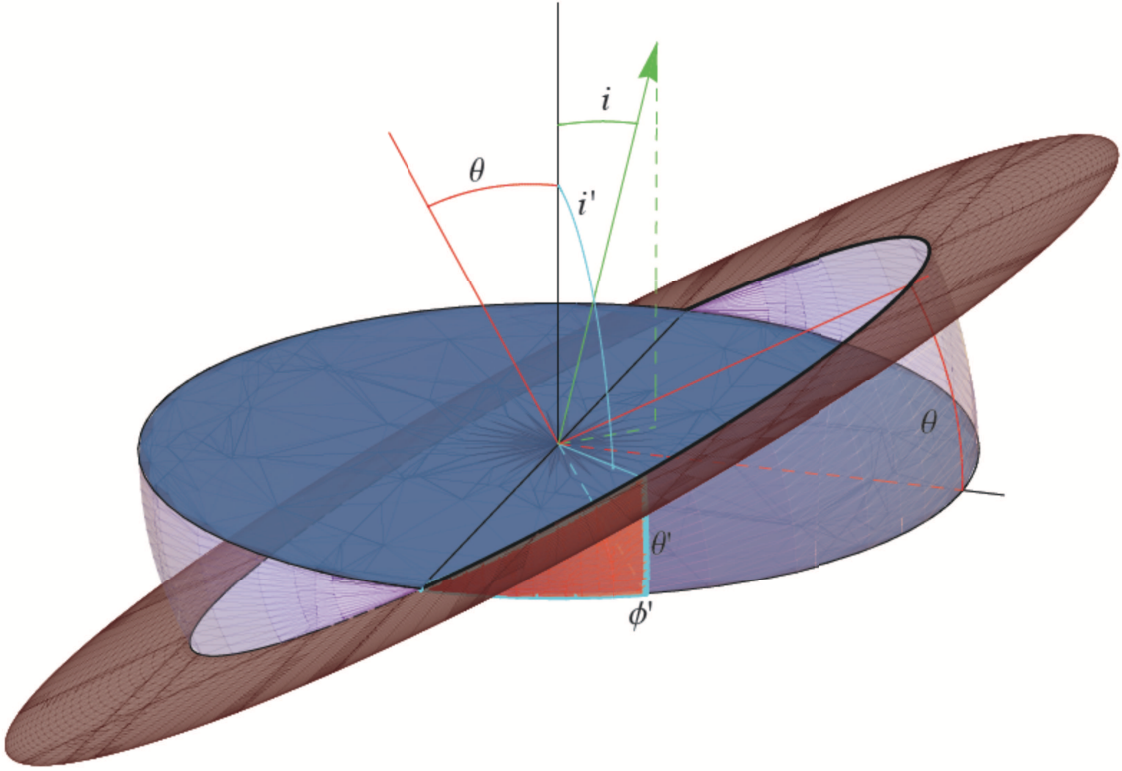} 
\caption{
Representation of a tilt-only disc (warped but not twisted). The outermost disc 
(shown here as an annulus) is tilted at an angle $\theta$ with respect to the 
inner disc. 
That is also the angle between the spin axes of the annulus and the inner disc. 
The transition from outer disc to inner disc occurs over a range of radii in reality, 
but is shown here happening at a single radius for convenience. Adopting the 
azimuth of the ascending node as $\phi=0$, at azimuth $\phi'$ the obscuration 
from the outer disc extends an angle $\theta'$ above the inner disc given by 
$\sin(\theta')=\sin(\theta)\sin(\phi')$.
The equivalent polar angle $i'  = \frac{\pi}{2} - \theta'$ is given by $\cos(i') = \sin (\theta) \sin(\phi)$.  
At each azimuth, the light purple shading represents the obscuration due to the 
tilted disc (which takes the form of two wedges, each of maximum width $\theta$). 
The light shadow represents the obscuration due to the tilted disc at all azimuths. 
The region at  azimuths $0\leq \phi \leq \phi'$ has a darker shadow to emphasize 
the angles intervening in the $\theta'$ calculation. 
}
\label{Fig_TiltedDisk-11Nov30.v2}
\end{figure*}  

\section{Summary and Conclusions} 
\label{sec:Conclusions}

In this paper, we combined an improved version of the MC97 disc wind model 
(Hall et al. 2012, in preparation) with the MHD driving of EBS92. 
We analysed how the resulting line profiles depend on different parameters of 
the model. In particular, we studied how the observing angle $i$ and the wind 
launch angle $\vartheta_0$ affect the emission line profiles. 
We found that for fixed $\vartheta_0$ all profiles are slightly asymmetric, with 
more asymmetric profiles for smaller inclination. 
For a given launching angle, less inclined objects have a larger fraction of their 
flux corresponding to motion towards the observer, therefore their profiles are 
less symmetric. 
For fixed $i$, the angle to be considered is $\vartheta > \vartheta_0$ , i.e the 
angle at which a wind that started at $\vartheta_0$ intercepts the base of the 
emitting region. Two different cases can be found. If $\vartheta > i$, wind 
velocity projections are mostly towards the observer, with red wings increasingly 
important for the cases $\vartheta \leq i$.
 
Our main conclusion is that the shape of the line profiles, their FWHMs and shift 
amounts (whether red or blue) with respect to the systemic velocity depend not 
only on the viewing angle but also on the angle (with respect to disc plane) at 
which the outflow starts. In fact, the relevant quantity is neither of the angles, but a 
combination of them. This is a consequence of how the model has been constructed. 
In effect, the optical depth expression includes a dependance on the wind contribution, 
which ranges from $v_{\rmn {p}} \sin(\vartheta + i)$ when $\phi = 0$, to  
$v_{\rmn {p}} \sin(\vartheta - i)$ when $\phi= \pi$. 
The launch angle parameter, although  included in the models, has been 
less explored in the literature. 
Note, however, that MC97 have reported that their \ionCiv \, line profiles do not 
strongly depend on $\vartheta_0$ ($\lambda_0$ in their notation).   
This difference could be due to our use of EBS92 streamlines instead of MC97 
streamlines, or it could be due to our more rigorous calculation of $L_\nu$ as 
compared to MC97. Similar results opposite to our findings are reported by 
\citet{Flohic+12}, in their study of Balmer emission lines. 

Using as a constraint the observational results obtained by \citet{Fine+10} for the 
\ionCiv \, lines in their sample, we found that the allowed region in the $i-\vartheta_0$ 
plane has an upper limit that depends on the torus half-opening angle, 
$i_{\rmn {max}}$. For instance, a launch angle $\vartheta_0  \sim 7\degr$ is only 
allowed for torus half-opening angle $\lesssim 20\degr$, while $\vartheta_0$ is 
$\lesssim 25\degr$ for $i_{\rm max} \sim 40\degr$. 
We found that the maximum torus half-opening torus angle that is compatible with the 
observations is about $47\degr$.   
Considering a model that replaces the torus with a tilt-only warped disc, formed by 
the alignment at smaller distances of material falling at large distances from random 
directions, yields no difference in the resulting allowed region of the inclination-launch 
angle plane. 

These results were obtained for a single mass and luminosity, as opposed to the 
\citet{Fine+10} and \citet{Mor+09} results which were obtained from datasets 
spanning an order of magnitude in both parameters. 
However, as mentioned in section \ref{sec:FWHMs}, Figure 2 of \citet{Fine+10} indicates 
a mild dependence of the dispersion of the linewidth on both these parameters. Thus, 
we expect that simulations for different masses and luminosities (currently being undertaken)
will yield similar results to those reported here. 
Future work will also consider applying the model to other high ionization lines, such as 
Si\,{\sc iv}, as well as low ionization lines, such as \ionMgii.

Other properties of the observed line profiles that could be measured 
and compared against the model include the line asymmetry \citep[e.g.][EBS92]{Whittle85a} 
and the cuspiness at $x$, $C_x$, related to the kurtosis and proposed by EBS92, where $x$ 
is defined as a given fraction of the peak flux. 
In their analysis of Balmer lines, \citet{Flohic+12} have also studied other line profile moments. 
In addition to the FWHM, they considered the full width at quarter maximum (FWQM) 
as well as the asymmetry and kurtosis indexes (A.I. and K.I. respectively), and centroid shift 
at quarter maximum ($v_{\rmn {c}}(1/4)$) as defined by Marziani et al. (1996). 
The present work does not include the treatment of resonance scattering of continuum 
photons or general relativistic (GR) effects.  
In their improvement of the MC97 and \citet{CM96} models, \citet{Flohic+12} found 
that when relativistic effects are included, the line profiles become skewed to the red. 
For a given combination of parameters, the line wings and centroids are increasingly 
redshifted with decreasing inclination. The amount of redshift is also a decreasing 
function of the inner radius of the line-emitting region, that satisfies 
$r_{\rmn{min}} \ge 100 \, r_{\rmn{g}}$, where $r_{\rmn{g}}$ is the gravitational radius. 
Although their results correspond to Balmer (i.e., low ionization) lines, we expect that high 
ionization lines such as \ionCiv \; 
may exhibit a similar or stronger response if the GR effects were to be considered,
as \ionCiv \;  is expected to be emitted predominantly at smaller radii.
However, our analysis and that of  \citet{Flohic+12} utilize different velocity fields,
and therefore a comparison of our results and theirs is perhaps not straightforward.
\citet{Giustini&Proga12} treat the absorption line dependence on wind 
geometry but we have not found in the literature a similar study for emission lines.
The relativistic MHD case has been studied by several authors, often in the context of jet launching 
and collimation and  in relation to several different astrophysical environments, such as AGN, 
microquasars, young stellar objects, pulsars and gamma-ray burst. 
The problem has been considered both in the steady \citep[e.g.][]{Camenzind86, Chiueh+91,  
Li+92, Contopoulos94, Contopoulos95, Fendt&Greiner01, Heyvaerts&Norman03} and in the 
time-dependent regimes \citep[e.g.][]{Koide+99, Porth&Fendt10}. 
In the relativistic MHD framework, the line formation problem has been considered in the X-ray 
range in relation to the iron K-line  \citep[e.g.][]{Muller&Camenzind04}. 
However, to the knowledge of the authors, the combination with a relativistic version of MC97 
has not been yet explored in the literature.  We consider that as one of the possible future lines 
of work to pursue.

\section*{Acknowledgments}
LSC and PBH acknowledge support from NSERC, and
LSC from the Faculty of Graduate Studies at York University.
PBH thanks the Aspen Center for Physics (NSF Grant \#1066293)
for its hospitality. 
The authors also would like to thank the anonymous referee for 
very helpful comments and suggestions.


\appendix

\section[]{Unobscured Sightline Distribution for Tilt-only Discs}
\label{app-sect:unobsc} 

Below we outline the evaluation of $dC(i)$ for arbitrary disc tilt angle $\theta$, to within 
the numerical factor required so that the total probability in a given situation is unity. 
Recall that in the following analysis the covering fraction is a function of the azimuth, 
while it is azimuthally integrated in the calculations of LE10.  

Define the line of nodes of the tilted outer disc relative to the inner disc to be at $\phi=0$. 
Then at each $\phi$, the obscuration from the outer disc extends an angle $\theta'$ above 
the inner disc given by $\sin\theta' = \sin (\theta) \sin(\phi)$ above the inner disc (sine rule 
for spherical right triangles); see Figure \ref{Fig_TiltedDisk-11Nov30.v2}.

The equivalent polar angle $i'  = \frac{\pi}{2} - \theta'$ is given by
$\cos(i') = \sin (\theta) \sin(\phi)$.
Solving the latter equation for the maximum unobscured $\phi$ at a given $i$,
$\phi_{\rmn {max}}(i,\theta)$, yields 
\begin{equation}
\phi_{\rmn {max}}(i,\theta)=\arcsin[\cos(i)/\sin\theta],  
\end{equation} 
where $0<\phi_{\rmn {max}}(i,\theta)<\frac{\pi}{2}$.

For $0<\theta<\frac{\pi}{2} $, there is an unobscured polar cap (at $i < \frac{\pi}{2} - \theta$)
and a region where obscuration increases from 0\% at $i=\frac{\pi}{2} -\theta$
to 50\% at $i=\frac{\pi}{2} $.  The differential solid angle in each region is:
\begin{align}
dC\left(i < \frac{\pi}{2} - \theta\right) &= 2 \pi \sin (i) di \label{eq: dC-iLessThanBeta} \\
dC\left(\frac{\pi}{2} - \theta < i < \frac{\pi}{2} \right) 
&= \sin(i) di \left(\pi + 2\int_{\phi=0}^{\phi_{\rmn {max}}(i,\theta)} d\phi\right) \nonumber \\
&= \sin (i) di \left(\pi + 2\arcsin\left[{\cos(i)\over \sin\theta}\right]\right) 
\label{eq: dC-iBetweenBetaAndPiOver2}
\end{align}
For $\frac{\pi}{2} <\theta<\pi$, there is a polar cap of complete obscuration 
$(i<\theta-\frac{\pi}{2} )$
and a region where obscuration decreases from 100\% at $i=\theta-\frac{\pi}{2} $ to 50\%
at $i=\frac{\pi}{2} $.  
The differential solid angle in the partially unobscured region is: 
\begin{equation} 
\begin{split}
dC\left(\frac{\pi}{2} + \theta < i < \pi\right) & = 
\sin (i) di \left(\pi - 2\int_{\phi=0}^{\phi_{\rmn {max}}(i,\theta)} d\phi\right)  \\ 
& = \sin (i) di \left(\pi - 2\arcsin\left[{\cos(i)\over \sin\theta}\right]\right) 
\end{split}
\end{equation}
For $0<\theta<\frac{\pi}{2} $, large $i$ values are underrepresented, while small $i$ 
values are underrepresented for $\frac{\pi}{2} <\theta<\pi$. 
At $\theta=\frac{\pi}{2}$, the half of the hemisphere with $0 < \phi < \pi$ is obscured, which 
leads to a uniform 50\% reduction in the probability of observing the quasar along every sightline 
as compared to the no-obscuration case. 

\subsection{Random orientations with $0 \leq \theta \leq \frac{\pi}{2}$.}

Here we analyse in more detail a restricted variant of the model 
where instead of a fixed tilt angle $\theta$, a distribution of such angles
randomly distributed in solid angle from $0 \leq \theta \leq \frac{\pi}{2}$ is considered.    
Combining equations \ref{eq: dC-iLessThanBeta} and \ref{eq: dC-iBetweenBetaAndPiOver2} 
we obtain: 
\begin{equation} 
\begin{split}
dC(i) & = 
\sin(i) di 
\Biggl[
\int_{\theta=0}^{\frac{\pi}{2}-i} 2 \pi \sin\theta d\theta \,+ \\ 
&  \int_{\theta = \frac{\pi}{2}-i}^{\frac{\pi}{2}} 
\left(\pi + 
2\arcsin\left[{\cos(i)\over \sin\theta}\right] \right) \sin\theta d\theta 
\Biggr], 
\end{split}
\end{equation}
which becomes 
\begin{equation} 
\begin{split}
dC(i) & = \sin(i) di 
\Biggl[
2\pi - \pi\cos\left(\frac{\pi}{2}-i\right) \\ 
& + 2 \int_{\theta=\frac{\pi}{2}-i}^{\frac{\pi}{2}} \arcsin\left[{\cos(i)\over \sin\theta}\right] 
\sin\theta d\theta 
\Biggr]
\end{split} 
\end{equation}

From $dC(i)$ we define $P(i)$, the probability of being unobscured, as 
\begin{equation} 
\label{eq:Prob-Unobsc--LE10-tilt-only-ThetaMax90}
P(i) = \frac{dC(i)}{2 \pi \sin(i)}.
\end{equation}  

In the text, 
we comment on the results of combining the latter expression with equations 
\ref{eq: mean-f} and \ref{eq:disp-f} to perform the same analysis applied in the Fine 
at al. case.

\bsp

\label{lastpage}

\end{document}